\documentclass[a4paper, 12pt, twoside]{article}

\usepackage[a4paper, margin=1in]{geometry}
\usepackage[utf8]{inputenc}
\usepackage[T1]{fontenc}
\usepackage[english]{babel}
\usepackage{textcomp} 
\usepackage{graphicx}
\DeclareGraphicsExtensions{.jpg,.mps,.pdf,.png,.gif}
\usepackage[french]{varioref}
\usepackage[]{amsmath,amssymb,amsfonts}

\usepackage{float}
\usepackage{bbm}
\usepackage{amsthm}
\usepackage{dsfont}
\usepackage{color}
\usepackage{natbib}
\usepackage{pict2e}
\usepackage{gensymb}
\usepackage{subcaption}
\usepackage[hyperindex,breaklinks]{hyperref}

\newcommand{\Ben}{\begin{enumerate}}
\newcommand{\Een}{\end{enumerate}}
\newcommand{\Bit}{\begin{itemize}}
\newcommand{\Eit}{\end{itemize}}
\newcommand{\Beq}{\begin{equation}}
\newcommand{\Eeq}{\end{equation}}
\newcommand{\Ba}{\begin{align*}}
\newcommand{\Ea}{\end{align*}}

\newcommand*\samethanks[1][\value{footnote}]{\footnotemark[#1]}
\newcommand{\Tb}{\textcolor{black}}

\title{Trends in the extremes of environments associated with severe US thunderstorms}
\date{October 30, 2019}

\begin{document}

\author{Erwan Koch\thanks{These authors contributed equally to this paper.} \thanks{Institute of Mathematics, EPFL, Station 8, 1015 Lausanne, Switzerland. \newline Emails: erwan.koch@epfl.ch, jonathan.koh@epfl.ch, anthony.davison@epfl.ch} ~~Jonathan Koh\samethanks[1] \samethanks[2] ~~Anthony C. Davison\samethanks[2] ~~ \\ Chiara Lepore\footnote{Lamont-Doherty Earth Observatory, Columbia University, Palisades, New York, USA. \newline Email: clepore@ldeo.columbia.edu} ~~ Michael K. Tippett\footnote{Department of Applied Physics and Applied Mathematics, Columbia University, New York, New York, USA. \newline Email: mkt14@columbia.edu}} 

\maketitle

\begin{abstract}
Severe thunderstorms can have devastating impacts. Concurrently high values of convective available potential energy (CAPE) and storm relative helicity (SRH) are known to be conducive to severe weather, so high values of PROD=$\sqrt{\mathrm{CAPE}} \times$SRH have been used to indicate high risk of severe thunderstorms. We consider the extreme values of these three variables for a large area of the contiguous US over the period 1979--2015, and use extreme-value theory and a multiple testing procedure to show that there is a significant time trend in the extremes for PROD maxima in April, May and August, for CAPE maxima in April, May and June, and for maxima of SRH in April and May. These observed increases in CAPE are also relevant for rainfall extremes and are expected in a warmer climate, but have not previously been reported. Moreover, we show that the El Ni\~no-Southern Oscillation explains variation in the extremes of PROD and SRH in February. Our results suggest that the risk \Tb{from severe thunderstorms} in April and May is increasing in parts of the US where it was already high, and that \Tb{the risk from storms} in February tends to be higher over the main part of the region during La Ni\~na years. \Tb{Our results differ  from those obtained in earlier studies using extreme-value techniques to analyze a quantity similar to PROD.}

\medskip

\end{abstract}

\section{Introduction}

Annual losses from severe thunderstorms in the US have exceeded \$10 billion in recent years.\footnote{\url{http://www.willisre.com/Media_Room/Press_Releases_(Browse_All)/2017/WillisRe_Impact_of_ENSO_on_US_Tornado_and_Hail_frequencies_Final.pdf}} In addition to economic losses, 2011 was marked by 552 deaths caused by tornadoes.  These economic and human impacts are a strong motivation to study how and why US thunderstorm activity varies from year to year and region to region. Two important aspects are trends potentially related to climate change or multi-decadal variability, and modulation by the El Ni\~no-Southern Oscillation (ENSO).  However, inadequacies in the length and quality of the thunderstorm data record present substantial challenges to addressing these questions directly \citep{Verbout2006,Allen:Hail:2015,Edwards2018:wind}.

\Tb{
In the US, a severe thunderstorm is defined to be one that produces a tornado, hail greater than one inch in diameter, or wind gusts in excess of 50 kts.
Supercell storms are responsible for a large fraction of severe thunderstorm reports (e.g., 79\% of tornadoes according to \cite{trapp2005tornadoes}), even though only about 10\% of thunderstorms are supercells \citep{Doswell2015Meso}, and a key element in forecasting severe thunderstorms is the prediction of where and when supercells will occur \citep{corfidi2017severe}. A supercell is a thunderstorm with a deep, long-lived rotating updraft (mesocyclone). The presence of buoyancy, i.e., convective available potential energy (CAPE), and deep-layer vertical wind shear are important determinants for supercell development. In addition to the magnitude of the vertical shear, the angle between surface and upper-level winds is important for mesocyclone development and persistence. A key quantity is atmospheric helicity, which is computed relative to storm motion and is proportional to vertical wind shear and the amount of wind direction turning from the surface to upper levels (often 0--3 km). 
}

Several recent studies of US tornado reports have concluded that annual numbers of reliably observed tornadoes, i.e., those rated E/EF1 and greater, show slight but statistically insignificant trends downward over time \citep{BrooksScience2014,Elsner:Tornado:efficiency:2014}, whereas measures of tornado outbreaks or clusters show upward trends \citep{BrooksScience2014,Elsner:Tornado:efficiency:2014,TippettCohen:ExtremeTornado}.  Changes in regional tornado activity have also been reported \citep{Agee2016tornado,Gensini2018}, but there is less evidence for changes in hail and damaging straight-line wind, perhaps due to the poorer quality of the relevant databases.

In view of the limitations of the historical storm record, a valuable alternative is the analysis of meteorological environments associated with severe thunderstorms. \Tb{As mentioned above, severe thunderstorms, especially supercell storms, are more likely in the presence of high values of CAPE and of certain measures of} vertical wind shear \citep[see, e.g.,][]{Brooks2003, brooks2013severe} such as storm relative helicity (SRH). Weather forecasters have routinely used such quantities for two decades to interpret observations and the output of numerical weather prediction models \citep{Johns1993,Rasmussen1998,Doswell1996}, and they are also useful in climatological studies, especially in areas outside the US without extensive historical reports \citep{Brooks2003}. The environmental approach can also provide an indication of expected severe thunderstorm activity in a warmer climate based on climate projections that do not resolve thunderstorms explicitly \citep{Trapp2009,Diffenbaugh2013}. On time-scales between weather forecasts and climate projections, this approach has provided a clearer picture of how ENSO modulates US hail and tornado activity \citep{Allen:ENSO2014,Lepore:ENSO:2017}.

However, there are notable gaps in previous statistical studies of environments associated with severe thunderstorms. For instance, relationships with ENSO were diagnosed based on \textit{monthly} averages, which are at best indirect proxies for behaviour on the time-scale of weather. Similarly \citet{Gensini2018} computed monthly accumulations of daily maxima of a significant tornado parameter. \citet{TippettCohen:ExtremeTornado} used submonthly environmental data but aggregated the results on an annual and US-wide basis. These gaps motivate the present work, which focuses on extremes of the environmental values rather than on monthly averages, and presents results that are spatially and temporally resolved. The framework that we use is statistical extreme-value theory.

\Tb{\cite{gilleland2013spatial} apply the conditional extreme-value framework of \cite{heffernan2004conditional} to the product WS$\times W_{\rm max}$, where WS is a measure of wind shear and $W_{\rm max}=\sqrt{2 \times \mathrm{CAPE}}$, by conditioning on the $75$th percentile of that variable computed across the spatial domain. This approach has the advantage of allowing the study of real spatial patterns under severe conditions, as opposed to approaches looking at pointwise maxima. They show some temporal variations in the mean simulated values from their model.}

\Tb{\cite{mannshardt2013extremes} perform an unconditional univariate analysis in which they fit the generalized extreme-value (GEV) distribution to} the annual maxima of WS$\times W_{\rm max}$ and establish the existence of a time trend in the GEV location parameter.  \Tb{\cite{heaton2011spatio} consider three Bayesian hierarchical extreme-value models based on exceedances over a high threshold for WS$\times W_{\rm max}$, their third model being based on a Poisson point process with a yearly time trend. Neither paper} clarifies whether this trend is attributable to both CAPE and WS or only to one of them. Moreover, both articles \Tb{consider trends in annual quantities} and thus cannot detect month-specific features, and they do not account for multiple testing, though this \Tb{issue} is briefly addressed in \cite{gilleland2008large}. Finally, they consider only time as a covariate. 

We propose to \Tb{address} some of the gaps left by the \Tb{papers mentioned in the previous paragraph}. Our study covers a large part of the contiguous US for individual months from 1979 to 2015. We separately consider CAPE, SRH (0--3 km) and the combined variable PROD=$\sqrt{\mathrm{CAPE}} \times$SRH.
\Tb{To motivate our use of PROD, we consider the discriminant line defined in \citet[][Equation (1)]{Brooks2003}, which is one of the first thresholds used to distinguish  low and high likelihoods of severe thunderstorm occurrence using a function of CAPE and vertical shear. This equation can be rewritten as $\mathrm{S6} \times \mathrm{CAPE}^{0.62}= 18.60$, where S6 is the 0--6 km shear. Replacing S6 with 0--3 km SRH and approximating the power $0.62$ by $0.5$ leads to a discriminant line of the form $\mathrm{SRH} \times \sqrt{\mathrm{CAPE}}=c$, i.e., $\mathrm{PROD}=c$, where $c$ is a real constant, and shows that values of PROD can be expected to be indicative of high risk of severe thunderstorms. PROD has already been used as a proxy for severe thunderstorms in several studies \citep[e.g.,][]{TippettCohen:ExtremeTornado} and the plot of Figure~1 in \cite{Brooks2003} is little changed by replacing S6 with 0--3 km SRH (not shown). More generally, the product of CAPE and two shear-related variables (different or not), or equivalently its square root, is commonly used as an indicator of the likelihood of severe thunderstorm occurrence. For instance, the significant tornado parameter (STP) and the supercell composite parameter (SCP) involve the product of CAPE, S6 and 0--1 km SRH, and the product of CAPE, S6 and 0--3 km SRH, respectively \citep[e.g.,][]{thompson2003close}.}  

\Tb{To ensure the soundness of our results} we carefully check the \Tb{suitability} of the GEV and the use of time and ENSO as explanatory variables in its location parameter, and we \Tb{account} for multiple testing by implementing the false discovery rate procedure of \cite{benjamini1995controlling}. \Tb{As stated in \citet[][Section 1]{gilleland2013spatial}, in addition to studying PROD, it is insightful to consider its components separately. Furthermore, accounting for multiple testing is essential when performing many  simultaneous tests, as  highlighted by \citet[Section 4]{gilleland2013spatial}, though they do not apply an adjustment for it.}

We find a significant time trend in the location parameter of the GEV for PROD maxima in April, May and August (and to a lesser extent in June and December), in CAPE maxima in April, May and June (and to a lesser extent in August, November and January), and in SRH maxima in May (and to a lesser extent in April). \Tb{The trends in CAPE maxima are striking, because CAPE is expected to increase in a warming climate \citep{del2007will, van2009surface} and are relevant to rainfall extremes \citep{lepore2015temperature}, but have not  previously been observed over the US.} April and May are important months for PROD, as severe thunderstorms are frequent at this period. The corresponding time slope is positive in regions of the US \Tb{where severe thunderstorms are already common}, which may have implications for risk assessment and management. Our study also reveals that ENSO can explain variation in the location parameter of the GEV for PROD and SRH maxima in February. The corresponding slope is negative over most of the region we consider, possibly suggesting an increased \Tb{risk of high storm impacts} in February during La Ni\~na years. \Tb{Our results differ from those of \citet{heaton2011spatio}, \cite{mannshardt2013extremes} and \cite{gilleland2013spatial}, but are fairly consistent with those obtained by \cite{Gensini2018}, who inter alia consider the numbers of tornado reports.} 

The remainder of the paper is organized as follows. Section~\ref{Sec_Data} presents the data and a brief exploratory analysis.  We describe our statistical approach and demonstrate its relevance in Section~\ref{Sec_Methodology}. Section~\ref{Sec_Results} details our main results, and Section~\ref{Sec_Conclusion} summarises our findings and discusses them.

\section{Data and exploratory analysis}
\label{Sec_Data}

The data we investigate consist of 3-hourly time-series of 0-180 hPa convective potential energy (CAPE, Jkg$^{-1}$) and 0--3 km storm relative helicity (SRH, m$^2$s$^{-2}$) from 1 January 1979 at 00:00 to 31 December 2015 at 21:00. The region covered is a rectangle  over the contiguous US from~$-110^\circ$~to~$-80^\circ$~longitude and~$30^\circ$~to~$50^\circ$~latitude and the resolution is 1$^\circ$ longitude and 1$^\circ$ latitude. These data constitute a coarse version of reanalysis data from the North American Regional Reanalysis (NARR); the original resolution is 32 km longitude and 32 km latitude \citep[see, e.g.,][]{Mesinger2006:NARR}. The region contains 651 grid points, with no data available for 32 grid points over the sea or lakes. Using these time series, we build 3-hourly time series of PROD=$\sqrt{\mathrm{CAPE}} \times$SRH, measured in m$^3$s$^{-3}$.

As a physical covariate we use monthly values of the NINO 3.4 index (${}^{\circ}\mathrm{C}$)  from 1979 to 2015, taken from the ERSSTv5 data set available on the NOAA Climate Prediction Center website. 

\begin{figure}
\centering
    \includegraphics[width=0.99\textwidth]{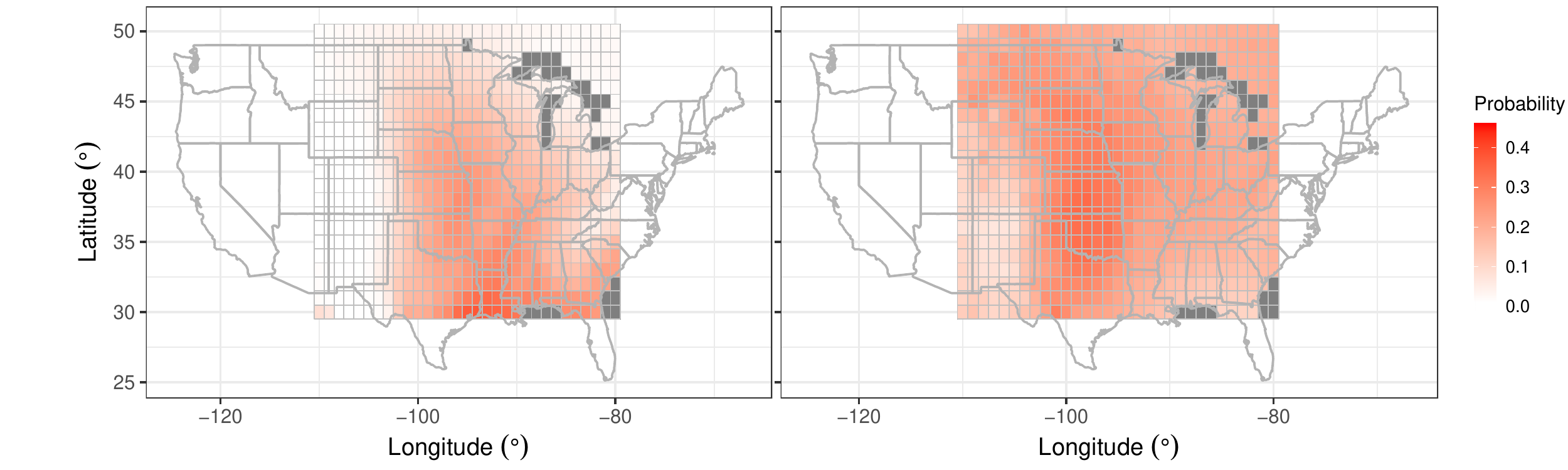} 
 \caption{Empirical pointwise probabilities of 3-hourly CAPE exceeding $1400$ Jkg$^{-1}$ (left) and SRH exceeding $170$ m$^2$s$^{-2}$ (right) for the entire period 1979--2015. Dark grey corresponds to grid points where no observations are available.}
 \label{figspatial}
\end{figure}

Figure~\ref{figspatial} shows the empirical pointwise probabilities that CAPE and SRH exceed thresholds corresponding to roughly the $90^{\text{th}}$ percentile of each variable across the entire region. There is a clear North-South gradient for CAPE probabilities, while the regional spatial pattern for SRH suggests that the high values cluster towards the centre of the region. 

\begin{figure}
\centering
    \includegraphics[width=.99\textwidth]{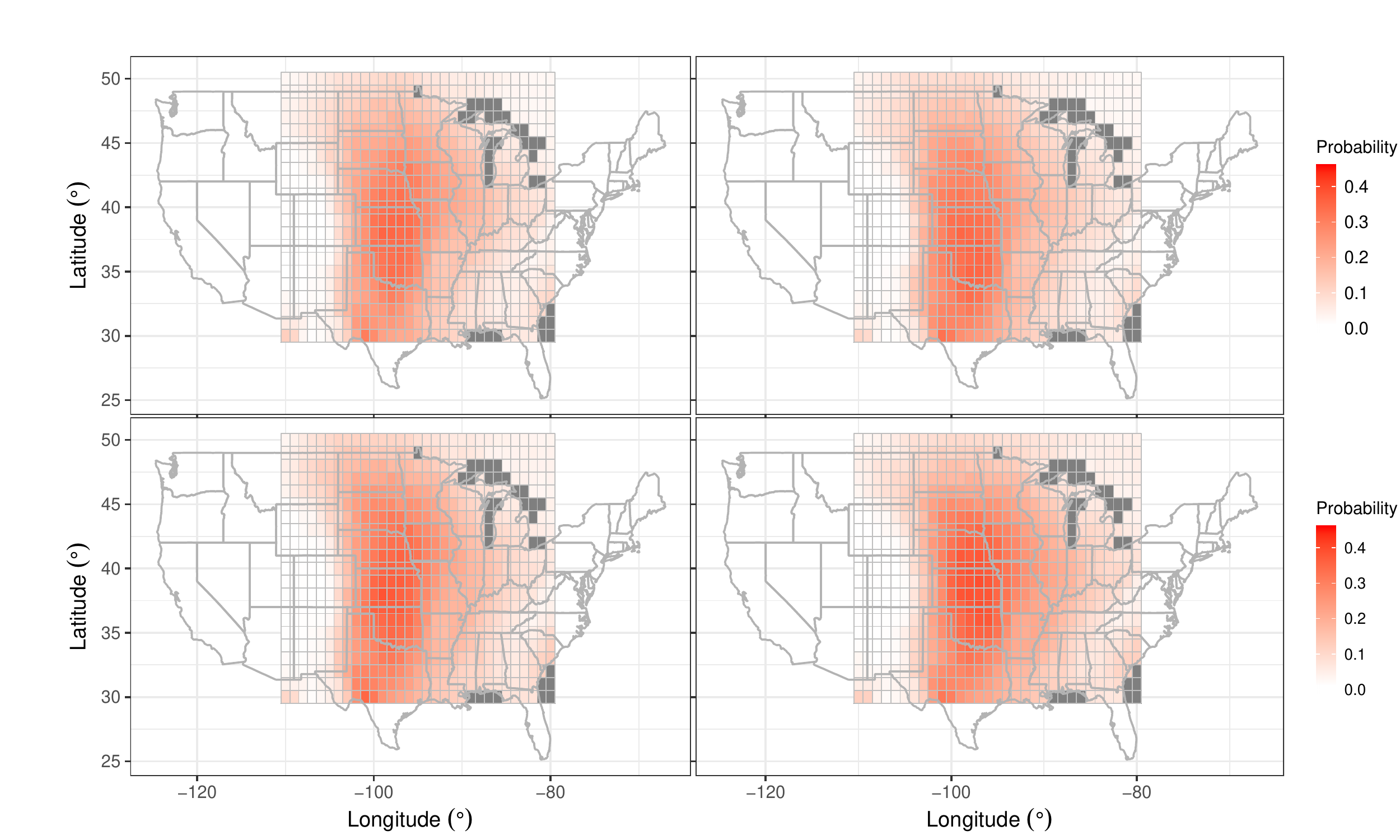}
  \caption{Empirical pointwise probabilities of 3-hourly PROD exceeding $3300$ m$^3$s$^{-3}$ during the periods 1979--1987 (top left), 1988--1996 (top right), 1997--2005 (bottom left) and 2006--2015 (bottom right).}
    \label{figtempprod}
\end{figure}

Figure~\ref{figtempprod} shows an increase in the exceedance probabilities for PROD at many grid points over the decades; a similar result is visible for SRH, but less so for CAPE. This increase is of interest for risk assessment, especially in regions with a high risk of severe thunderstorms. Figure~\ref{figtempprod} strongly suggests the presence of  a temporal trend in the maxima, but \Tb{there seems to be no geographical shift, notwithstanding  the results of \cite{gilleland2013spatial}}.

\begin{figure}
    \centering
      \begin{subfigure}[b]{\linewidth}
        \centering
    \includegraphics[width=.99\textwidth]{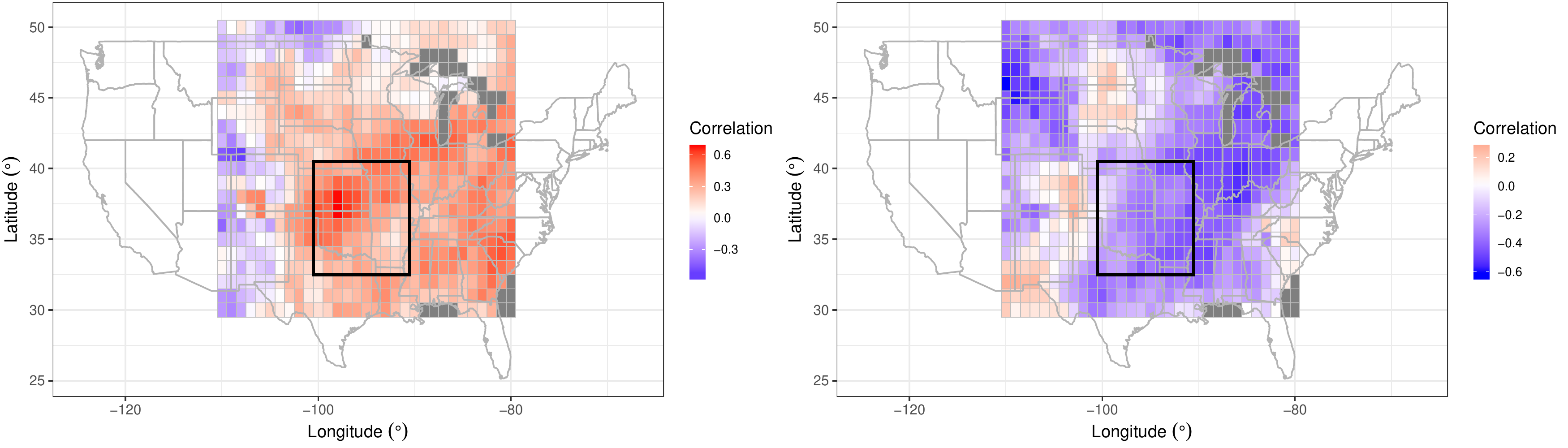}
  \end{subfigure} \\ 
  \vspace{4mm}
  \centering
  \begin{subfigure}[b]{.33\linewidth}
    \includegraphics[width=.99\textwidth]{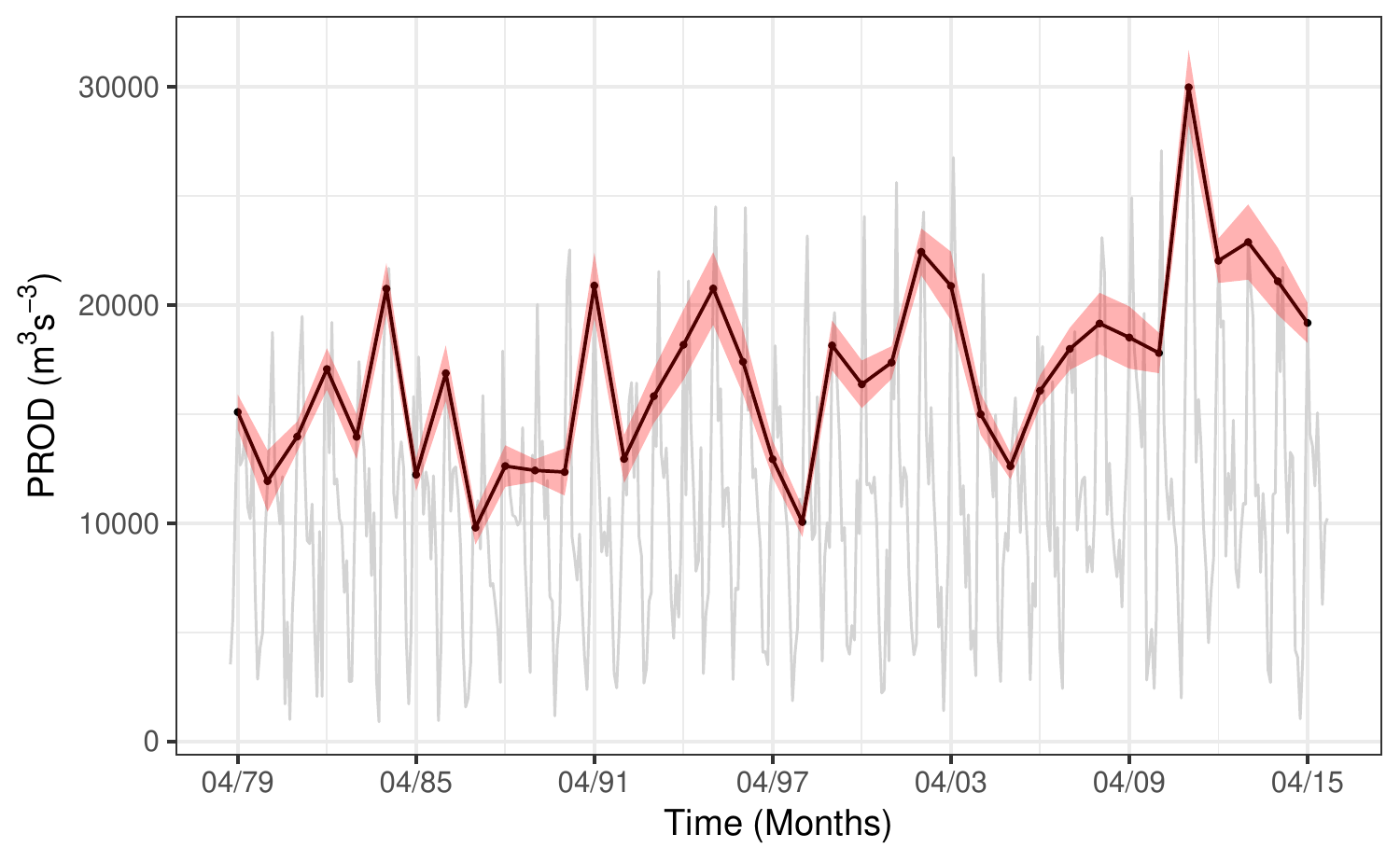}
  \end{subfigure}%
  \begin{subfigure}[b]{.33\linewidth}
    \centering
    \includegraphics[width=.99\textwidth]{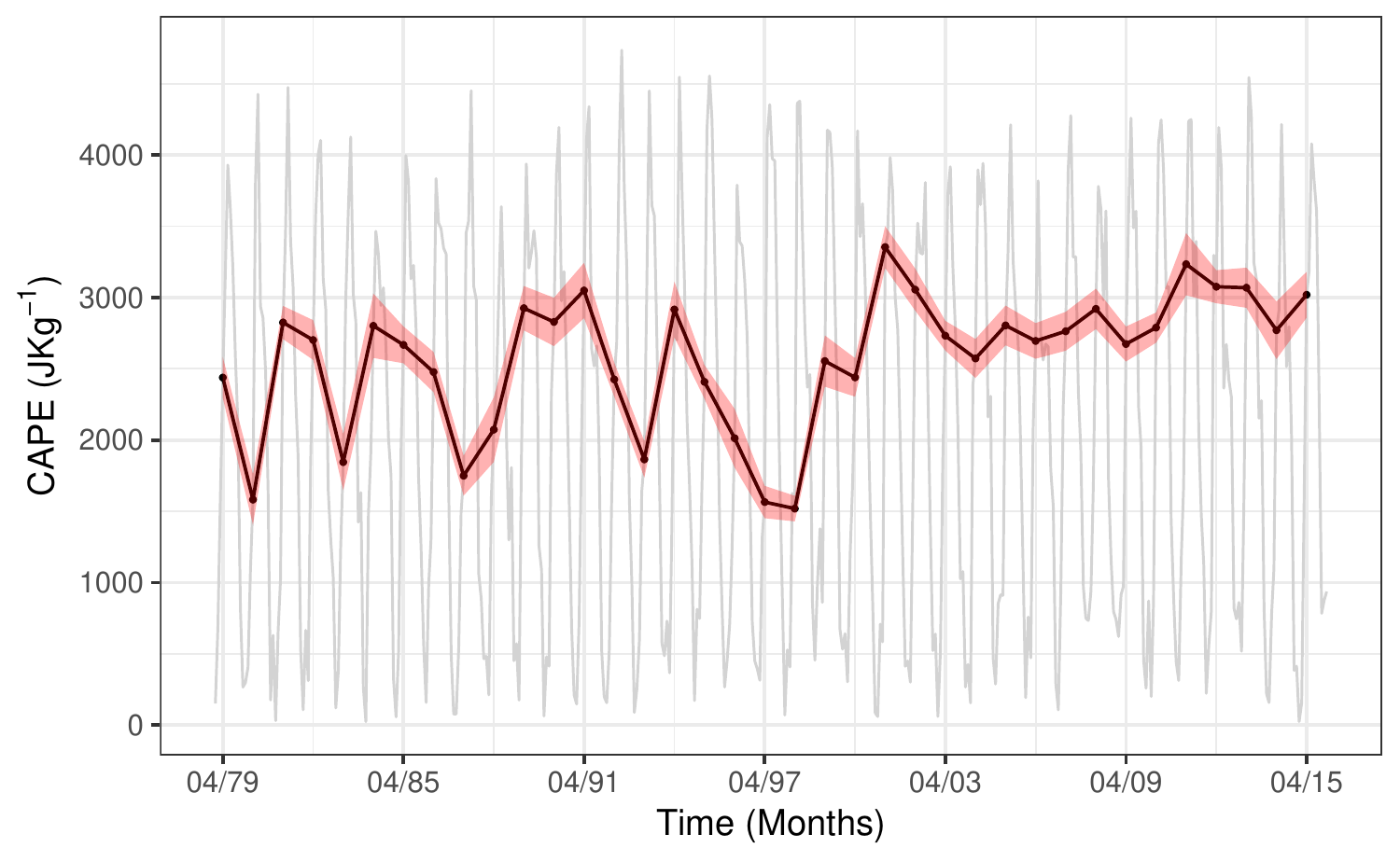}
  \end{subfigure} 
    \begin{subfigure}[b]{.33\linewidth}
    \centering
    \includegraphics[width=.99\textwidth]{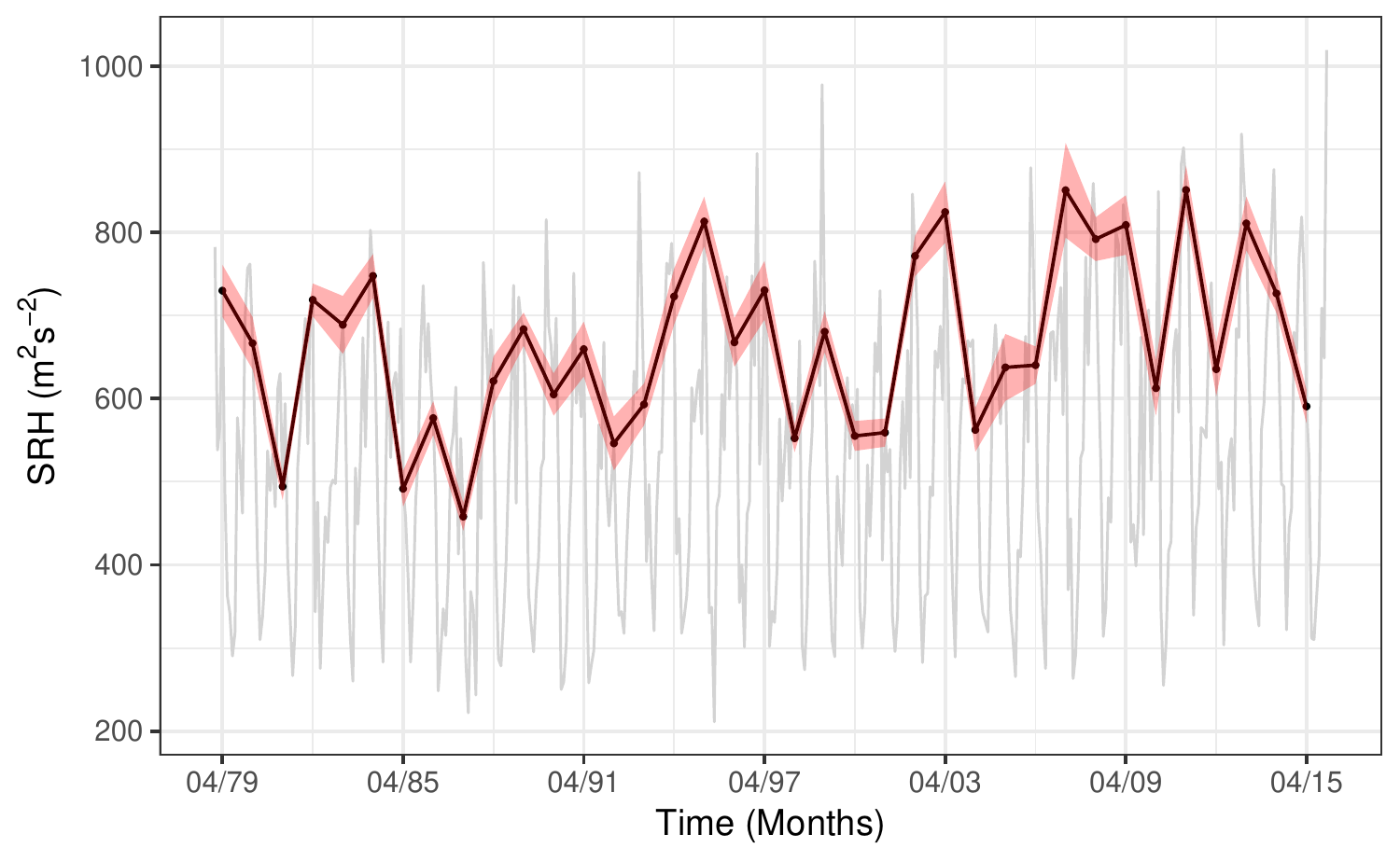}
  \end{subfigure} \\
    \vspace{4mm}
    \begin{subfigure}[b]{.33\linewidth}
    \centering
    \includegraphics[width=.99\textwidth]{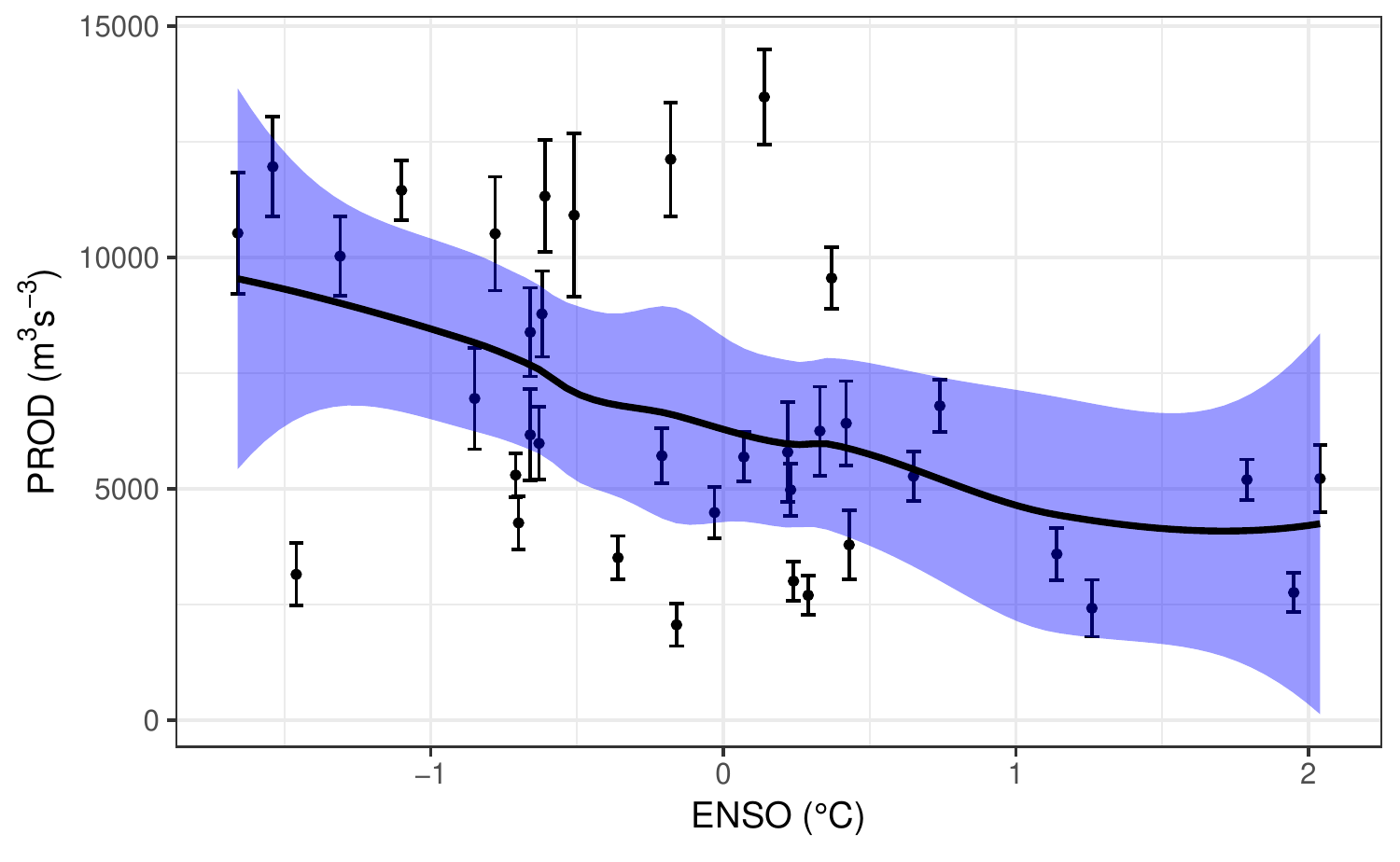}
  \end{subfigure}%
  \begin{subfigure}[b]{.33\linewidth}
    \centering
    \includegraphics[width=.99\textwidth]{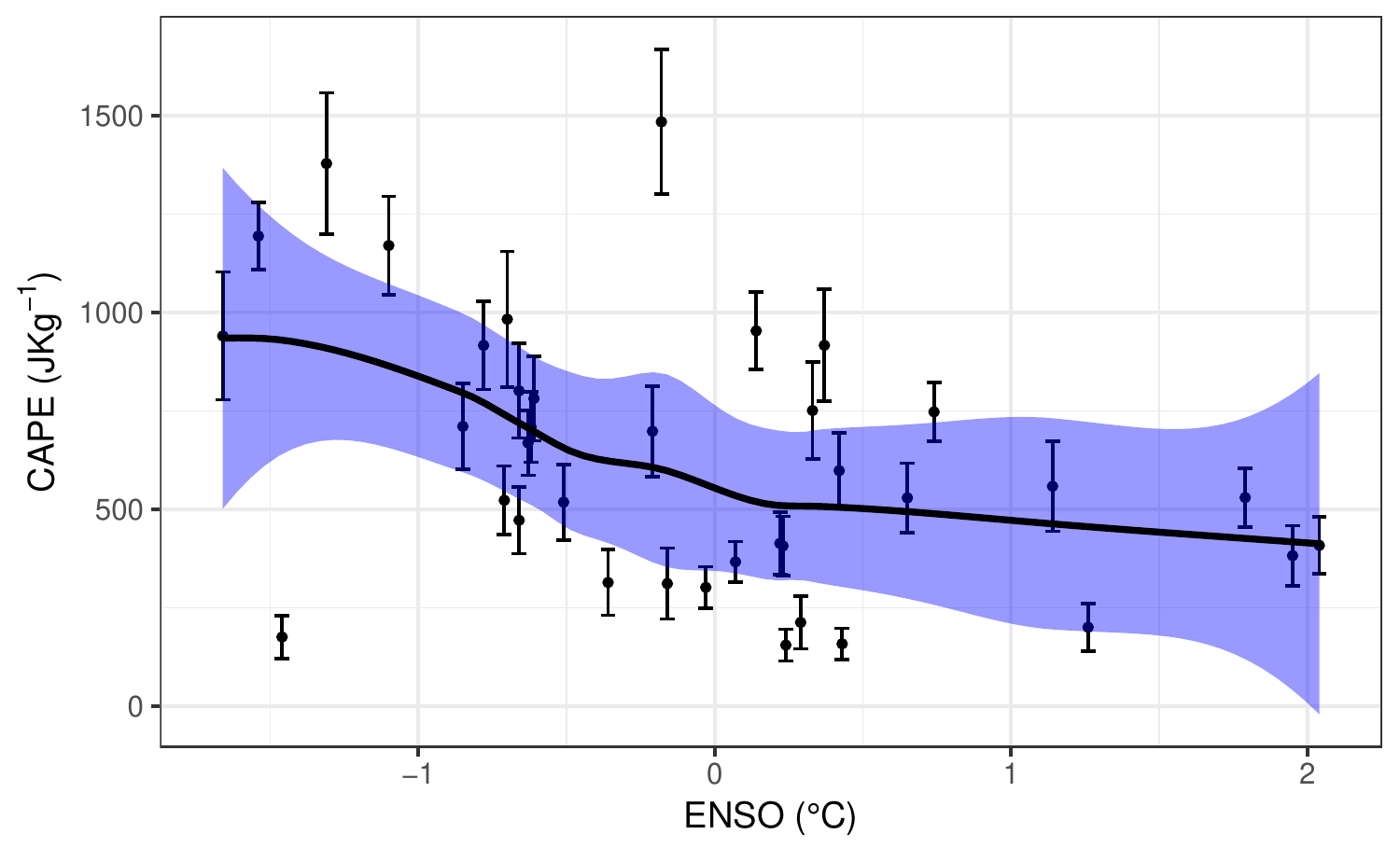}
  \end{subfigure} 
    \begin{subfigure}[b]{.33\linewidth}
    \centering
    \includegraphics[width=.99\textwidth]{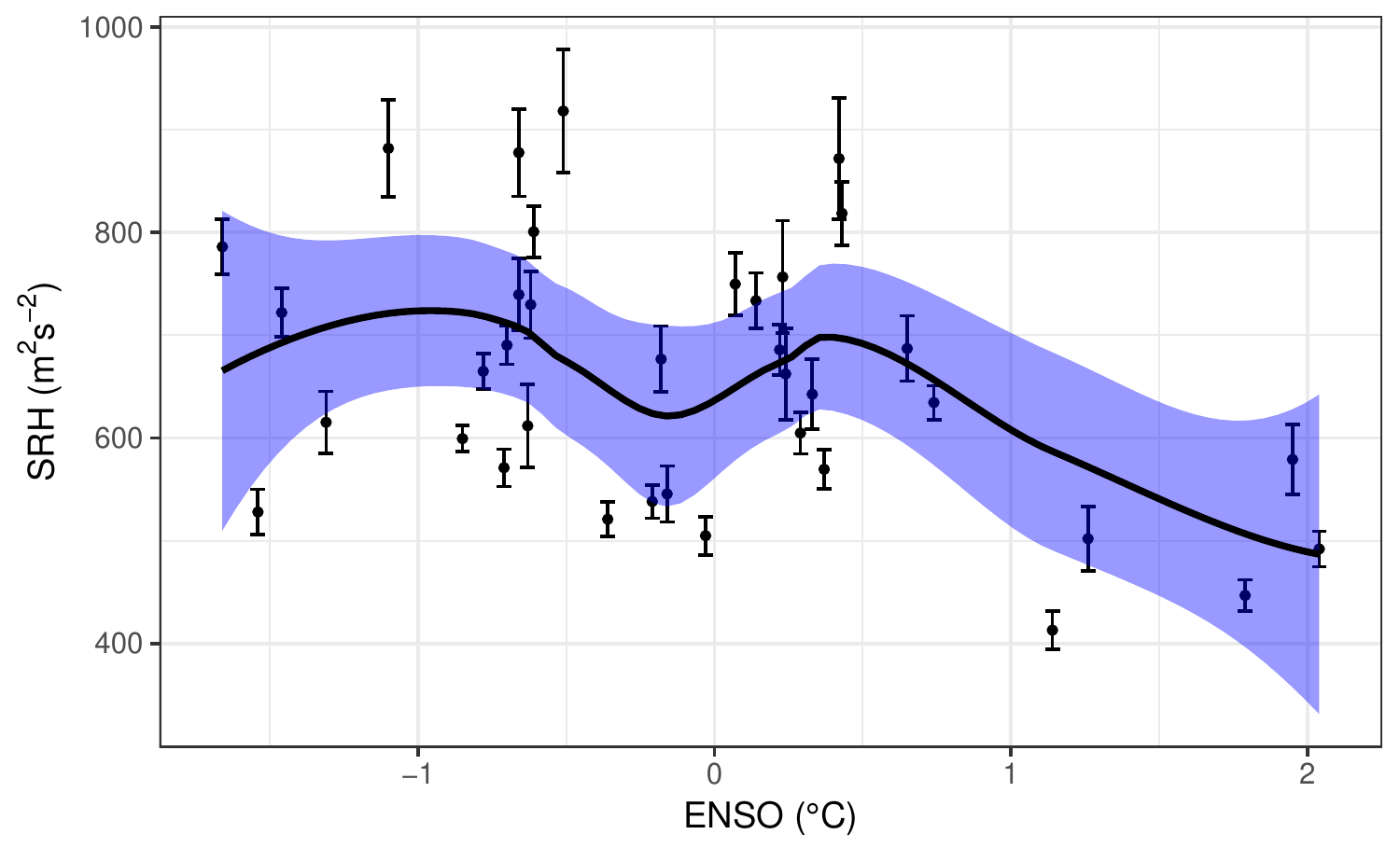}
  \end{subfigure} \\
    \caption{Exploratory analysis for monthly maxima: The top panels show the correlation map with time (in years from 1 to 37) for PROD April maxima (left) and the correlation map with ENSO for PROD February maxima (right). The middle and bottom panels display PROD (left), CAPE (centre) and SRH (right) analyses on a subregion indicated by the black rectangle drawn on the correlation maps. The middle panels show the region-averaged monthly maxima time series across all 444 months in light grey, the region-averaged April maxima time series in black and its $95\%$ confidence interval bounds indicated by the red shaded region. Every point in the time series is the averaged maxima across all grid points in the subregion indicated before, for a particular month and a particular year. The bottom panels show scatter-plots of the region-averaged February maxima with ENSO, along with the $95\%$ confidence interval bounds at each point indicated by the whiskers. The black line represents the best fitted local regression trend estimate, with its $95\%$ confidence interval bounds indicated by the shaded blue region.}
  \label{prodseasonts}
\end{figure}

The top left panel of Figure~\ref{prodseasonts} shows a positive correlation between PROD April maxima and time  
for many grid points, and the middle panels show a positive linear time trend for April maxima of PROD, CAPE and SRH in the subregion indicated. The top right panel shows strong negative correlation between PROD February maxima and ENSO at many grid points, while the scatter-plots in the bottom panels show a roughly linear negative trend for all variables. These analyses underscore the need to incorporate ENSO into our statistical modelling of maxima. 

\section{Methodology}
\label{Sec_Methodology}

\subsection{Modelling of maxima}
\label{Subsec_Model_Maxima}

Risk assessment entails the estimation of return levels associated with very high return periods and of the probabilities of observing events so extreme that they have never occurred before. Extreme-value theory provides a solid framework for the extrapolation needed to perform these tasks for the maxima of PROD, CAPE and SRH. Here we present the statistical background to the results in Section~\ref{Sec_Results}; for further explanation and references see \citet{Coles:2001} or \citet{Davison.Huser:2015}.

Let $M_n$ denote the maximum of the independent and identically distributed random variables $X_1, \dots, X_n$. 
The extremal types theorem states that if there exist sequences $\{ a_n \}>0$ and $\{ b_n \} \in \mathbb{R}$ such that $(M_n-b_n)/a_n$ has a non-degenerate limiting distribution as $n\to\infty$, then this must be a generalized extreme-value (GEV) distribution, 
$$
\mathrm{GEV}_{\eta, \tau, \xi}(x)=\left \{
\begin{array}{ll}
\exp \left[ - \left \{ 1+\xi (x-\eta)/\tau \right \}^{-1/\xi}_{+} \right] ,& \quad \xi \neq 0, \\
\exp \left[ -\exp \left \{ -(x-\eta)/\tau \right \}_{+} \right ], & \quad \xi =0, 
\end{array}
\quad x \in \mathbb{R},
\right.
$$
where $\xi$ and $\eta$ are real-valued, $\tau>0$ and, for any real $a$, $a_+=\max \{a,0 \}$. This implies that if $n$ is large enough, we may approximate the distribution of $M_n$ by 
\Beq
\label{Eq_Distr_Maxima}
\mathbb{P}(M_n \leq x) \approx \mathrm{GEV}_{\eta, \tau, \xi}(x), \quad x \in \mathbb{R},
\Eeq
for suitably chosen $\eta$, $\tau$ and $\xi$, which are location, scale and shape parameters. 
The latter defines the type of the distribution: $\xi>0$,  $\xi <0$ and $\xi=0$ correspond to the Fr\'echet, Weibull and Gumbel types and allow quite different statistical behaviours, with the first giving a heavy upper tail with polynomial decay, the second modelling bounded variables, and the third an intermediate case, unbounded with an exponentially-decaying upper tail. 

The GEV approximation for maxima remains valid if the variables are dependent, provided that distant extremes are ``nearly independent'' (more formally, Leadbetter's $D(u_n)$ condition is satisfied).  We shall see below that this appears to be the case for our time series, so it is plausible that ~\eqref{Eq_Distr_Maxima} applies.
 
\begin{figure}
    \centering
      \begin{subfigure}[b]{.33\linewidth}
    \centering
    \includegraphics[width=.99\textwidth]{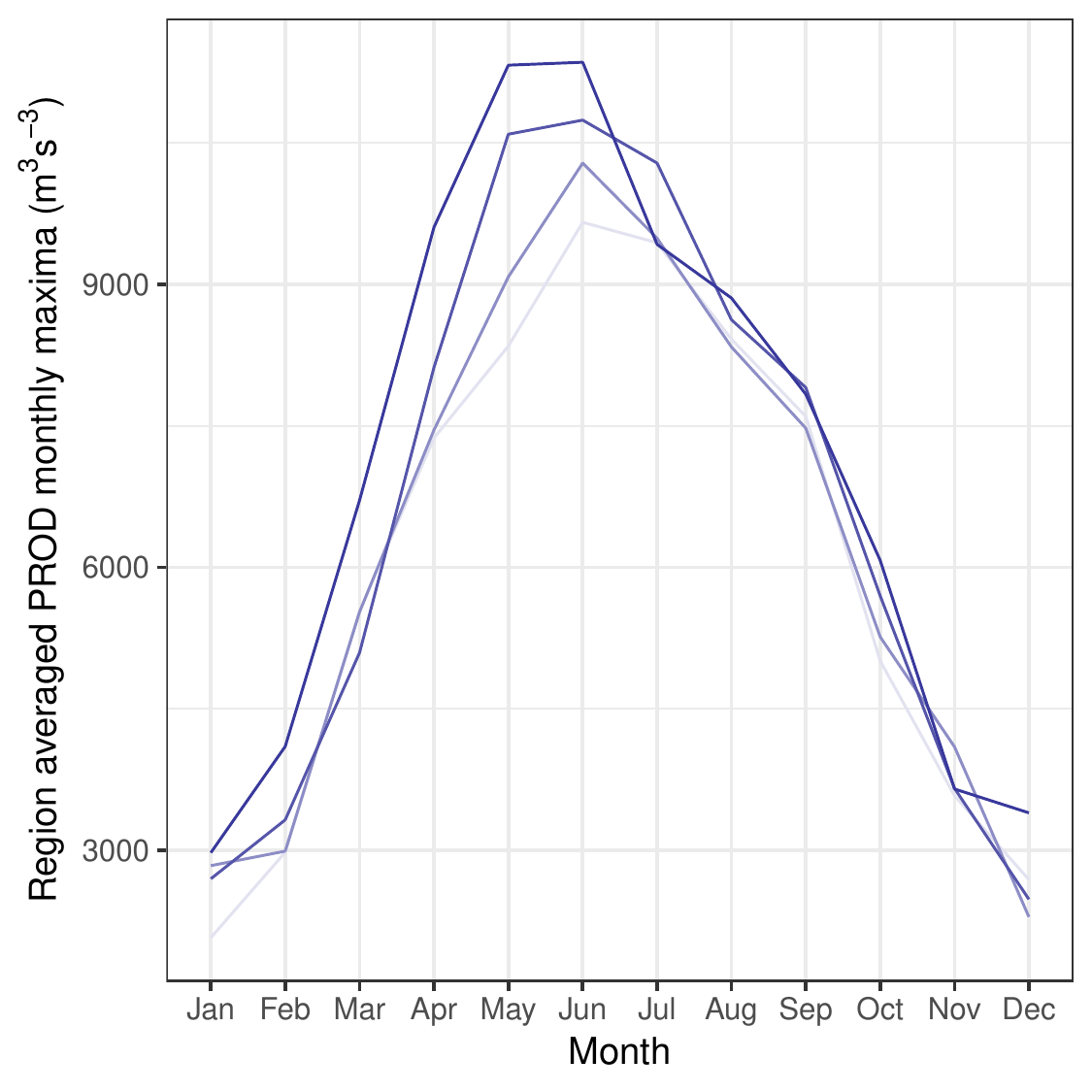}
  \end{subfigure}%
  \begin{subfigure}[b]{.33\linewidth}
    \centering
    \includegraphics[width=.99\textwidth]{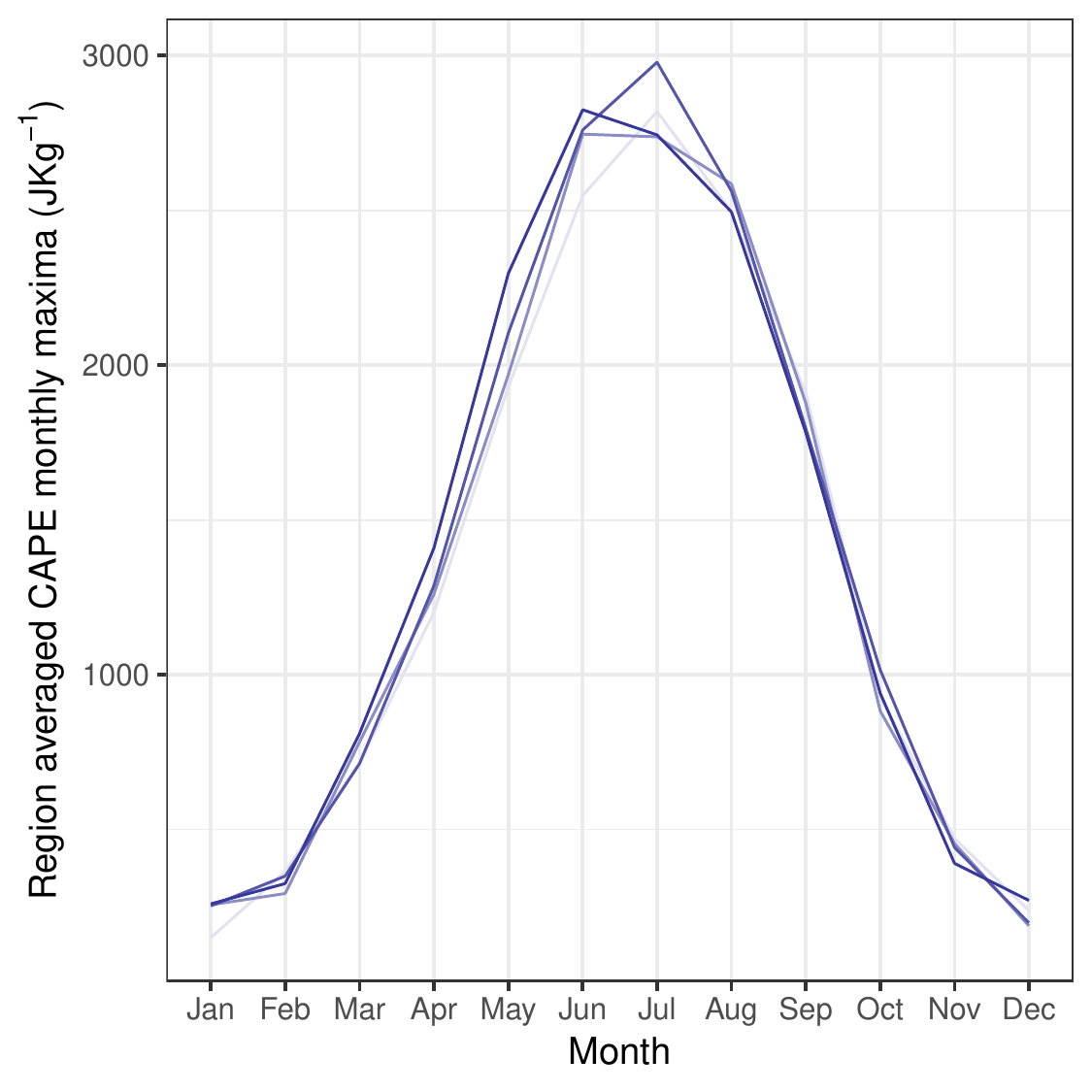}
  \end{subfigure}
  \begin{subfigure}[b]{.33\linewidth}
    \centering
    \includegraphics[width=.99\textwidth]{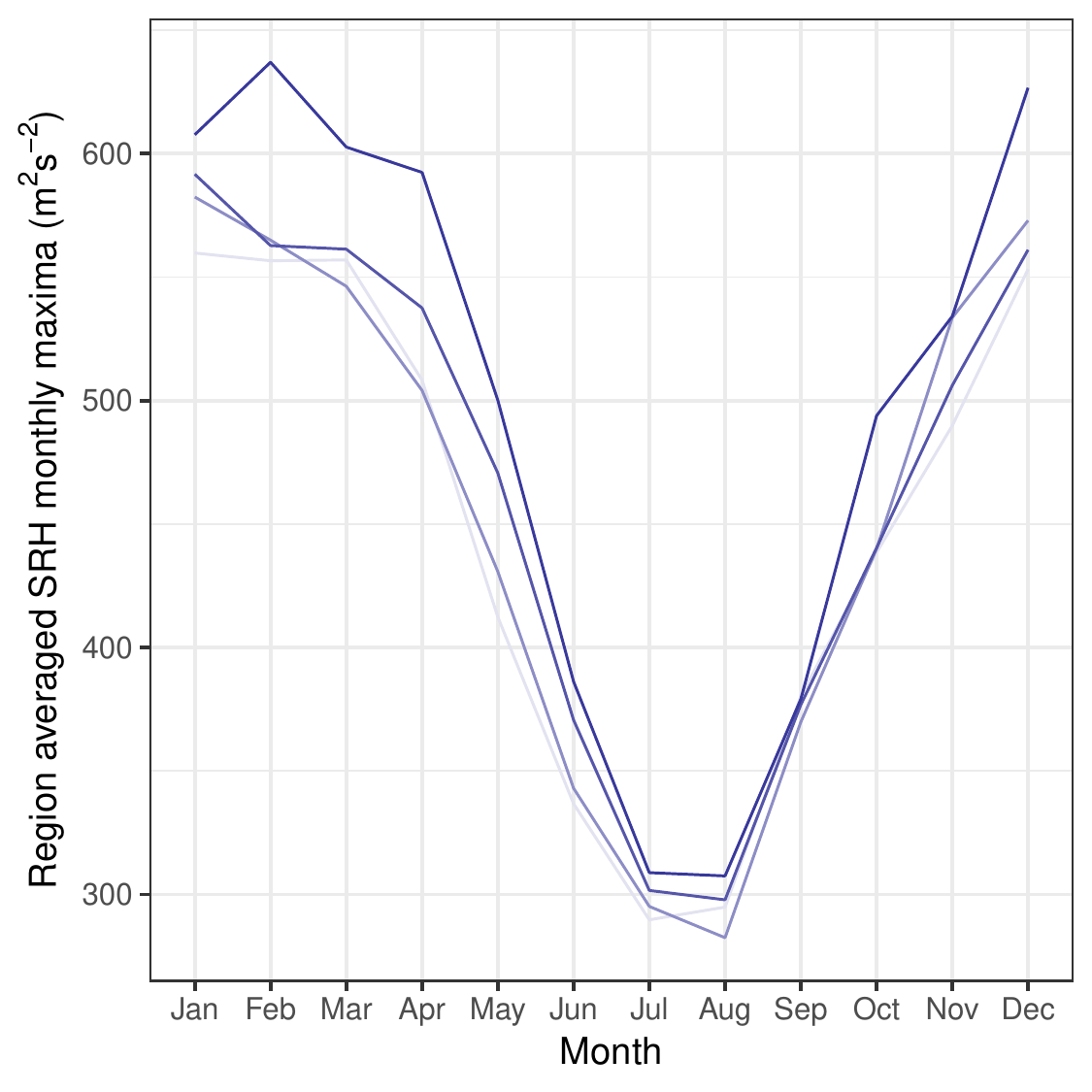}
  \end{subfigure}\\ 
  \caption{Whole region-averaged monthly maxima of PROD (left), CAPE (centre) and SRH (right). The four lines coloured from light blue to dark blue correspond to the time periods 1979--1987, 1988--1996, 1997--2005 and 2006--2015, respectively.}
  \label{figseason}
\end{figure}

The results above provide a natural model for maxima of stationary sequences. To apply this model we split the data into blocks of equal lengths and compute the maximum of each block. Assume that we have $T$ blocks of length $n$ and let  $M_n^{(1)}, \dots, M_n^{(T)}$ denote the corresponding maxima. If $n$ is large enough, the distribution of the $M_n^{(t)}$ is approximately~\eqref{Eq_Distr_Maxima}, upon which inference can be based; this is the so-called block maximum method. As noted in Section~\ref{Sec_Data}, PROD, CAPE and SRH maxima exhibit a time trend and/or a relation with ENSO for some months, and we can allow the GEV parameters to depend upon these variables. Figure~\ref{figseason} and results in Section~\ref{Sec_Results} show that the temporal or ENSO effects only appear for certain months. For instance, time trends for PROD, CAPE and SRH are mainly present in April and May, April to June and April and May, respectively. We therefore choose our blocks to be the months and study each month separately, fitting the models 
\begin{equation}
\label{Eq_Parameters_GEV_Function_Time}
M_n^{(t)} \sim \mathrm{GEV}_{\eta_{\mathrm{ti}}(t), \tau_{\mathrm{ti}}, \xi_{\mathrm{ti}}}, \quad \eta_{\mathrm{ti}}(t)=\eta_{0, \mathrm{ti}} + \eta_{1, \mathrm{ti}} t,  \quad t=1, \dots, T,
\end{equation}
and
\begin{equation}
\label{Eq_Parameters_GEV_Function_ENSO}
M_n^{(t)} \sim \mathrm{GEV}_{\eta_{\mathrm{en}}(t), \tau_{\mathrm{en}}, \xi_{\mathrm{en}}}, \quad \eta_{\mathrm{en}}(t)=\eta_{0, \mathrm{en}} + \eta_{1, \mathrm{en}} \mathrm{ENSO}_t, \quad t=1, \dots, T,
\end{equation}
where $\eta_{0, \mathrm{ti}}$, $\eta_{1, \mathrm{ti}}$, $\eta_{0, \mathrm{en}}$, $\eta_{1, \mathrm{en}}$, $\xi_{\mathrm{ti}}$ and  $\xi_{\mathrm{en}}$ are real-valued, $\tau_{\mathrm{ti}}$ and $\tau_{\mathrm{en}}$ are positive, ${\rm ENSO}_t$ is the value of ENSO in that month for year $t$, and $n$ equals 224, 232, 240 or 248, depending on the number of days in the month, as we have eight observations per day.  Figure~\ref{prodseasonts} suggests that effects of time and ENSO on maxima are roughly linear and impact the location parameter $\eta$ only, so we consider constant scale and shape parameters; it is generally inappropriate to allow the \Tb{shape parameter} depend on a covariate owing to the large uncertainty of its estimate. The time trend induces non-stationarity between the blocks (i.e., across years) but does not violate the within-block stationarity assumption; see below. Figure~\ref{figseason} suggests that the time trend does not stem from a shift of seasonality.

We compute the monthly maximum for each month and a given grid point and thereby obtain the maxima $M_{31}^{(1)}, \dots, M_{31}^{(37)}$ for January, say.  We then fit the models~\eqref{Eq_Parameters_GEV_Function_Time} and~\eqref{Eq_Parameters_GEV_Function_ENSO} by numerical maximum likelihood estimation for each month and grid point.

Recall that, \Tb{provided the block size $n$ is large enough}, within-block stationarity and the $D(u_n)$ condition ensure the validity of~\eqref{Eq_Distr_Maxima} and hence allow us to consider the models~\eqref{Eq_Parameters_GEV_Function_Time} and~\eqref{Eq_Parameters_GEV_Function_ENSO}. To check the plausibility of these two properties, we considered the 3-hourly time series of PROD, CAPE and SRH at $50$ representative grid points. For each block (associated with a triplet grid point-month-year), we fitted several autoregressive-moving average (ARMA) processes to the corresponding time series, chose the fit that minimized the Akaike information criterion (AIC), and used a Box--Pierce procedure to assess the independence of the corresponding residuals; we found no systematic departure from independence or stationarity. Often the residual distribution appeared to lie in the Fr\'echet or Gumbel maximum-domains of attraction, and  \citet[Section 5.5]{Embrechts} show that in such cases convergence of the maxima to the GEV occurs even for ARMA processes. Hence the time series of data within the months seem to satisfy both stationarity and the $D(u_n)$ condition.  Choosing the months as blocks thus appears reasonable, as is confirmed by our analysis in \Tb{the following section}. \Tb{On the other hand, choosing the seasons or years as blocks would mask many interesting features, and the} sample size associated with day- or week-long blocks is too low for the GEV approximation~\eqref{Eq_Distr_Maxima} to be reasonable. 

\subsection{Assessment of GEV fit}
\label{Subsec_GEV_Test_Fit}

\begin{figure}
    \centering
      \begin{subfigure}[b]{.33\linewidth}
    \centering
    \includegraphics[width=.99\textwidth]{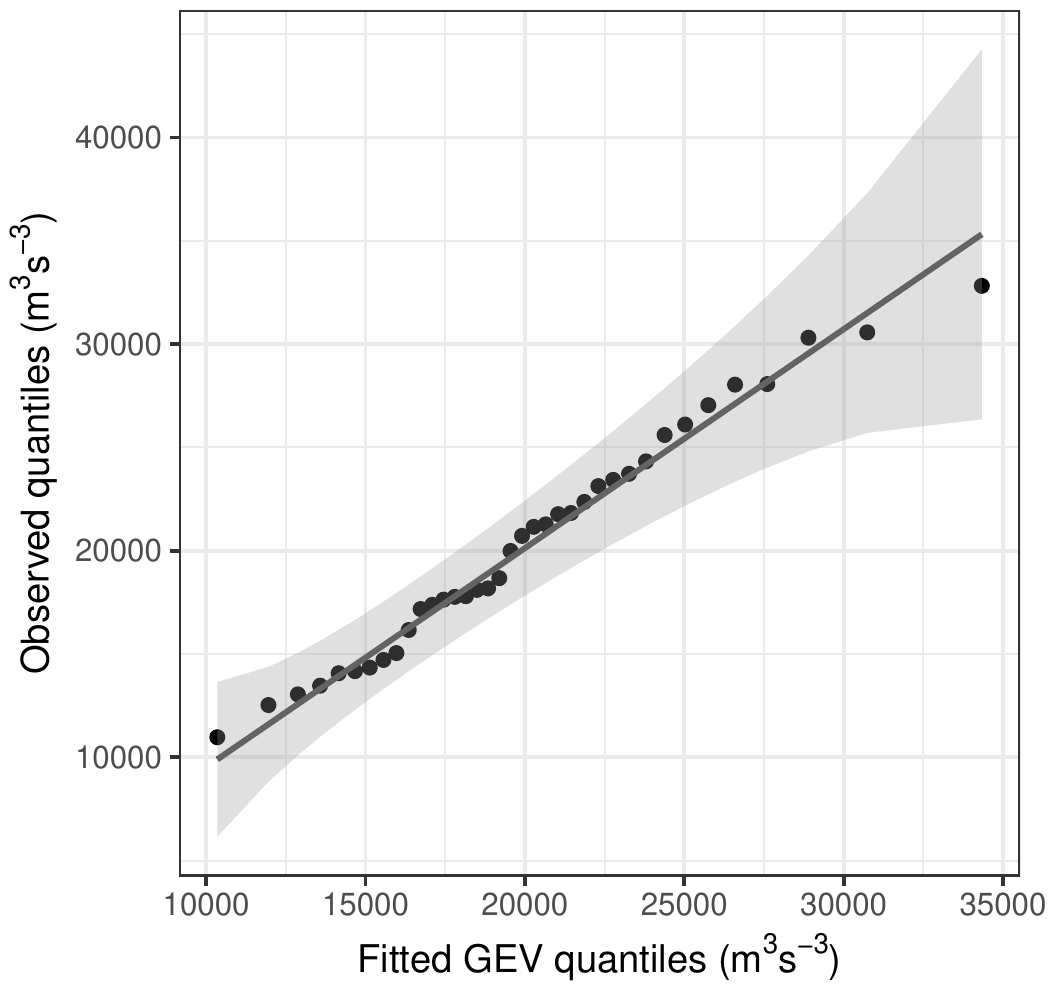}
  \end{subfigure}%
  \begin{subfigure}[b]{.33\linewidth}
    \centering
    \includegraphics[width=.99\textwidth]{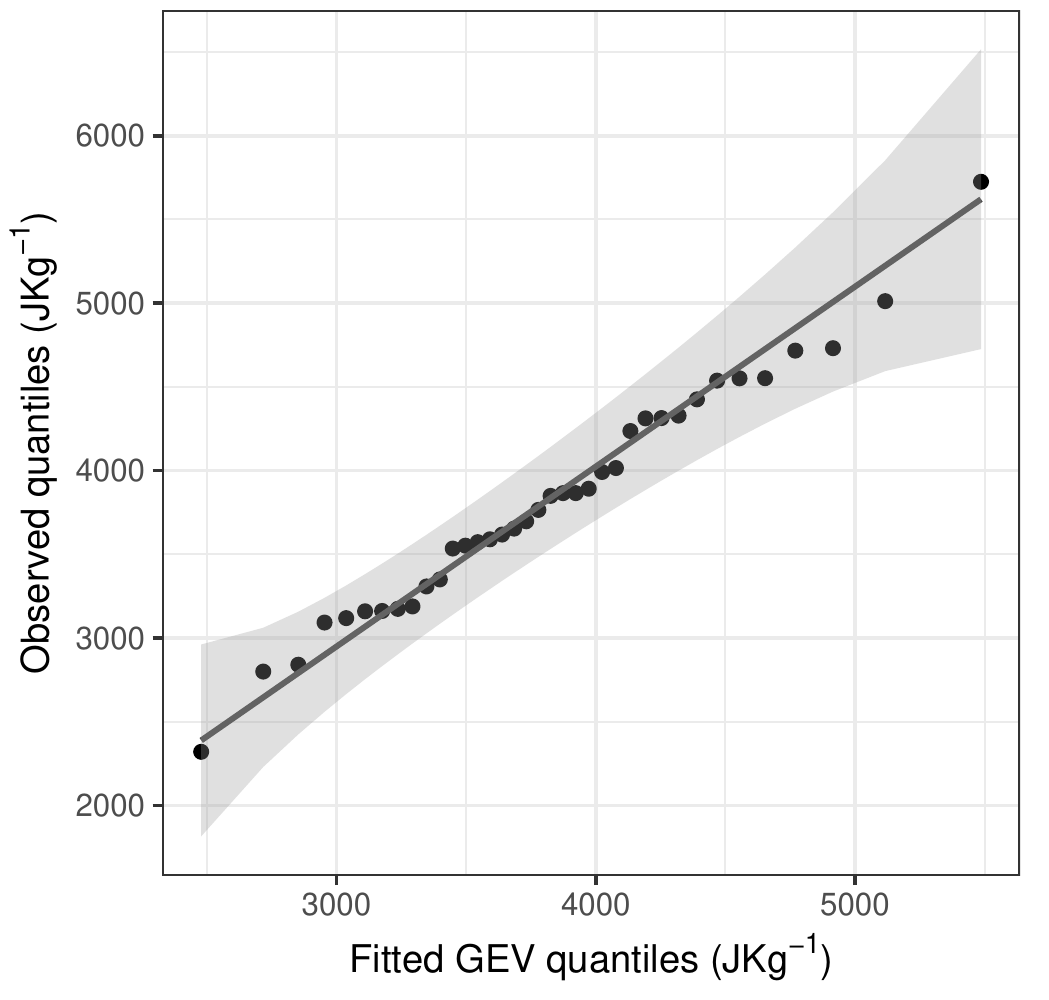}
  \end{subfigure}
  \begin{subfigure}[b]{.33\linewidth}
    \centering
    \includegraphics[width=.99\textwidth]{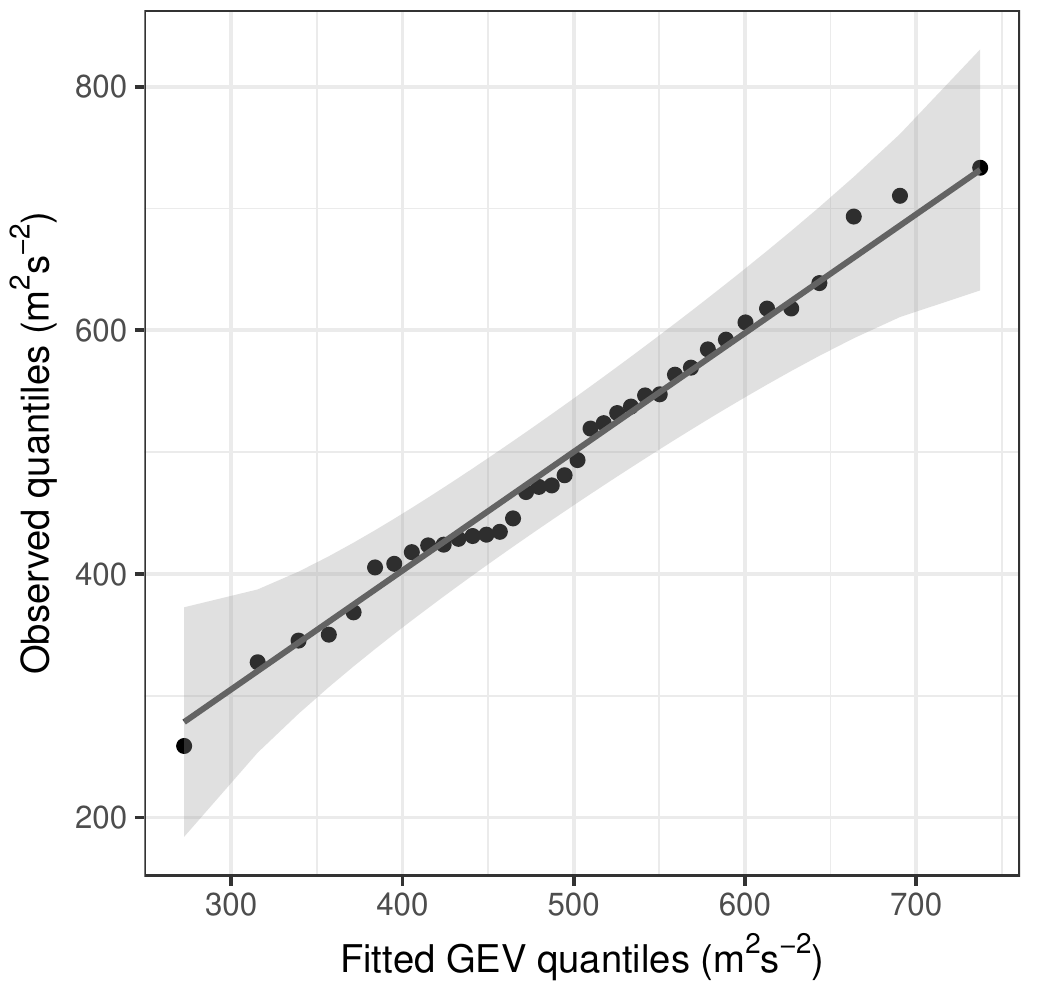}
  \end{subfigure}\\ 
  \caption{Assessment of the in-sample fit of the GEV: QQ plots for PROD (left), CAPE (centre) and SRH (right) May maxima at the grid point whose South-West corner has coordinates $32^\circ$ latitude and $-99^\circ$ longitude. The shaded regions indicate the $95\%$ confidence bounds.}
  \label{qqplot}
\end{figure}

At each grid point $i$ and month $j$, we fit the GEV to the monthly maxima, as described in \Tb{Section~\ref{Subsec_Model_Maxima}}, resulting in location, scale and shape parameter estimates $\hat{\eta}_{i,j}$, $\hat{\tau}_{i,j}$ and $\hat{\xi}_{i,j}$. We use the Kolmogorov--Smirnov test to assess the distributional proximity between this GEV and the empirical distribution of the $37$ observed monthly maxima. For PROD, CAPE and SRH, in most months, the fit appears acceptable at the $5\%$ level  at all grid points. These good in-sample fits of the GEV for all variables are confirmed by the quantile-quantile (QQ) plots, which are displayed for one grid point in Figure~\ref{qqplot}.

However, these results do not take into account the fitting of the GEV to the data, which systematically decreases the values of the Kolmogorov--Smirnov statistic. In order to make an informal allowance for this, we perform the following procedure for each grid point $i$ and month $j$:
\begin{enumerate}
\item fit the GEV using the pooled observations from the eight grid points nearest to $i$ to obtain $\hat{\eta}_{\mathrm{po}_i, j}$, $\hat{\tau}_{\mathrm{po}_i, j}$ and $\hat{\xi}_{\mathrm{po}_i, j}$;
 \item use a Kolmogorov--Smirnov test to check the agreement between the ``out-sample'' GEV with parameters $\hat{\eta}_{\mathrm{po}_i, j}$, $\hat{\tau}_{\mathrm{po}_i, j}$ and $\hat{\xi}_{\mathrm{po}_i, j}$, and the empirical distribution of the $37$ observed monthly maxima at grid point $i$. 
\end{enumerate}
Then, we implement the same procedure $100$ times with data simulated from independent GEV fitted at each grid point and compute the $5\%$ and $95\%$ quantiles of the empirical distribution of the number of rejections. Table~\ref{table:ks} shows that, for all variables, the observed numbers of rejections are low compared to the number of grid points (619), especially as we did not account for multiple testing. Moreover, they are not tremendously different from those obtained in the simulation study, \Tb{although often slightly above the $95\%$ quantile in the case of CAPE and slightly below the $5\%$ quantile for SRH and PROD}. These discrepancies may be explained by the substantial spatial dependence in our data, not accounted for in the simulation study. This procedure supports the use of the GEV at grid points at which no data are available and thus goes beyond the initial goal of assessment of the GEV fit.
\begin{table}
\centering
\renewcommand{\arraystretch}{1.5}
\resizebox{\textwidth}{!}{
\begin{tabular}{c||cccccccccccc}
 \textbf{Variable} &  \textbf{Jan} &  \textbf{Feb} &  \textbf{Mar} &  \textbf{Apr} &  \textbf{May} &  \textbf{Jun} &  \textbf{Jul} &  \textbf{Aug} &  \textbf{Sep} &  \textbf{Oct} &  \textbf{Nov} &  \textbf{Dec} \\ 
  \hline
      \hline
   \textbf{PROD}  & 42 & 27 & 50 & 29 & 52 & 58 & 67 & 71 & 55 & 27 & 39 & 34 \\ 
   \textbf{Sim PROD 5\%} & 37 & 34 & 33 & 33 & 35 & 35 & 37 & 38 & 38 & 36 & 36 & 37 \\  
   \textbf{Sim PROD 95\%} & 57 & 57 & 55 & 55 & 57 & 58 & 57 & 58 & 57 & 58 & 58 & 57 \\ 
   \hline \hline
   \textbf{CAPE}  & 59 & 33 & 48 & 34 & 60 & 64 & 71 & 90 & 67 & 43 & 58 & 74 \\ 
   \textbf{Sim CAPE 5\%} & 41 & 36 & 37 & 36 & 36 & 36 & 38 & 42 & 39 & 36 & 36 & 37 \\ 
   \textbf{Sim CAPE 95\%} & 59 & 58 & 60 & 56 & 57 & 58 & 61 & 63 & 61 & 60 & 58 & 59 \\   
   \hline \hline
   \textbf{SRH} & 36 & 23 & 24 & 21 & 22 & 42 & 42 & 34 & 35 & 26 & 24 & 36 \\ 
   \textbf{Sim SRH 5\%} & 36 & 36 & 34 & 35 & 34 & 36 & 36 & 36 & 34 & 36 & 36 & 34 \\ 
   \textbf{Sim SRH 95\%} & 60 & 59 & 59 & 53 & 57 & 57 & 57 & 58 & 58 & 58 & 61 & 57 \\ 
\end{tabular}
}
\caption{Assessment of the out-sample fit of the GEV: Number of rejections from our out-sample Kolmogorov-Smirnov test (at the $5\%$ level and without accounting for multiple testing) for each variable and each month. For each part (corresponding to one variable), the first row gives the observed number of rejections whereas the second and third ones provide the $5\%$ and $95\%$ quantiles of the empirical distributions of the number of rejections obtained from the simulation study.}
\label{table:ks}
\end{table}

We conclude that the GEV provides a suitable model for the monthly maxima of our three variables.

\subsection{Testing procedure}

\subsubsection{General}

In Section~\ref{Sec_Results}, we assess whether time and ENSO affect the location parameter of the fitted GEV for the three variables PROD, CAPE and SRH. However, as this is assessed at 619 grid points, we must make some allowance for multiple hypothesis testing. 

We first discuss the statistic used to test the significance of time and ENSO, respectively, in~\eqref{Eq_Parameters_GEV_Function_Time} and~\eqref{Eq_Parameters_GEV_Function_ENSO}. In the first case, we have to test the null and alternative hypotheses 
$$ 
H_0: \eta_{1, \mathrm{ti}}=0 \quad \mbox{versus} \quad H_A: \eta_{1, \mathrm{ti}} \neq 0,
$$
by comparing the fits of the models 
$$
\mathcal{M}_0: \eta_{\mathrm{ti}}(t)=\eta_{0, \mathrm{ti}} ,  \quad \mathcal{M}_1: \eta_{\mathrm{ti}}(t)=\eta_{0, \mathrm{ti}} + \eta_{1, \mathrm{ti}} t, \quad t=1, \dots, 37,
$$
and similarly for ENSO. We let $\ell_0(\mathcal{M}_0)$ and $\ell_1(\mathcal{M}_1)$ denote the maximized log-likelihoods for the models $\mathcal{M}_0$ and $\mathcal{M}_1$ and compute the signed likelihood ratio statistic $\tilde{T}=\mathrm{sgn}(\hat\eta_{1, \mathrm{ti}})[2 \{ \ell_1(\mathcal{M}_1)-\ell_0(\mathcal{M}_0) \}]^{1/2}$, where $\mathrm{sgn}(\hat\eta_{1, \mathrm{ti}})$ is the sign of the estimated trend under model $\mathcal{M}_1$; $\tilde T$ has an approximate standard Gaussian distribution under $H_0$, and the corresponding $p$-value is $p=2\Phi(-|\tilde t|)$, where $\tilde t$ is the observed value of $\tilde T$ and $\Phi$ denotes the standard Gaussian distribution function.  Computing $p$  for the  $m$ grid points yields $m$ ordered $p$-values $p_{(1)} \leq p_{(2)} \leq \dots \leq p_{(m)}$.  The underlying p-values are likely to be positively correlated, since dependence on time or ENSO will have a spatial component, and we now discuss how to adjust for this. 

\subsubsection{Multiple testing}
\label{Subsubsec_Multiple_Testing}

A popular approach for multiple testing in climatology is the field significance test of~\cite{livezey1983statistical}, but unfortunately this gives little insight about where the results are significant, which is of high interest to us, and the regression approach of~\cite{delsole2011field} has the same drawback. Among methods to identify grid points where the results are significant are those, such as the Bonferroni method, that strongly control the so-called family-wise error rate, i.e., the probability that the number of falsely rejected null hypotheses is equal or larger than unity. However, when the number of hypotheses to test is high, such methods  are so stringent that the power of the test is very low. 

\cite{benjamini1995controlling} introduce the false discovery rate (FDR), namely the expected proportion of false rejections  of the null hypothesis $H_0$ out of all rejections of it, and propose a procedure to ensure that the FDR is below a given level $q$ when performing multiple testing. Their approach, which we call the BH procedure, would reject $H_0$ at all grid points $i$ such that $p_i \leq p_{(k)}$, where
$$ 
k=\max \left \{ i: p_{(i)} \leq q \frac{i}{m} \right \}.
$$
In fact this ensures that the FDR is less than $q m_0/m$, where $m_0$ denotes the unknown number of grid points at which $H_0$ is true.  We then say that the procedure controls the false discovery rate at  level $qm_0/m$. 

For a chosen $q$, let $S_q$ be the number of grid points at which a particular covariate is declared significant by the BH procedure. Then we expect the true number of grid points where the relation is significant, $m_A$, to satisfy 
\Beq
\label{Eq_First_Lower_Bound_Number_Significance}
m_A \geq (1-q)S_q.
\Eeq
As the BH procedure ensures that the false discovery rate  is not more than $q m_0/m$, we may argue a posteriori that we have controlled the FDR at level 
$$
q^{(1)} = \frac{q\{m-(1-q)S_q\}}{m}\leq q{m_0\over m},
$$
which entails that $m_A \geq (1-q^{(1)})S_q$.
Iterating this argument by defining
$$
q^{(n+1)} = \frac{q \left\{ m- \left(1-q^{(n)} \right) S_q \right\}}{m}, \quad n =1,2,\ldots,
$$
the effective level at which we have controlled the FDR is therefore $q_{\lim}=\lim_{n \to \infty} q^{(n)}$. This limit is generally obtained after a few iterations. Finally, we may write that 
\Beq
\label{Eq_Lower_Bound_Number_Significance}
m_A \geq (1-q_{\lim}) S_q.
\Eeq
The BH procedure was originally shown to be valid for independent test statistics, but \citet[Theorem~2.1]{benjamini2001control} prove that it controls the FDR at level $q m_0/m$ if the statistics have a certain form of positive dependence. \cite{ventura2004controlling} apply the BH procedure to simulations representative of climatological data and covering the range of correlation scales likely to be encountered in practice, and find that it controls the FDR at level $qm_0/m$. \cite{yekutieli1999resampling} and \cite{benjamini2001control} propose modifications to account for more general dependence between the test statistics. The first is complicated and its gain over the BH procedure is limited, while the second is applicable whatever the dependence structure but has greatly reduced power, so \cite{ventura2004controlling} recommend the use of the BH procedure.

The independence assumption underlying the BH procedure is clearly false for our data, \Tb{but they resemble} those considered in \cite{ventura2004controlling}, so applying the BH procedure at level $q$ should control the FDR at level $q m_0/m$. A more rigorous argument would use the asymptotic normality of our test statistic $\tilde{T}$ and the multivariate central limit theorem to show that $\tilde{T}_1, \ldots, \tilde T_m$ is asymptotically jointly Gaussian and that the results of \citet{benjamini2001control} can be applied, but this is outside the scope of the present paper. 

\section{Results}
\label{Sec_Results}

In this section we quantify the effects of time and ENSO in the location parameter of the GEV and study their significance, using $q=0.05$ and $q=0.2$, corresponding to control of the false discovery rate at the nominal levels 5\% and 20\%. In each case we first discuss PROD, which is the main variable of interest for severe thunderstorm risk, and then consider  CAPE and SRH.  

\begin{figure}
\center
\includegraphics[scale=0.67]{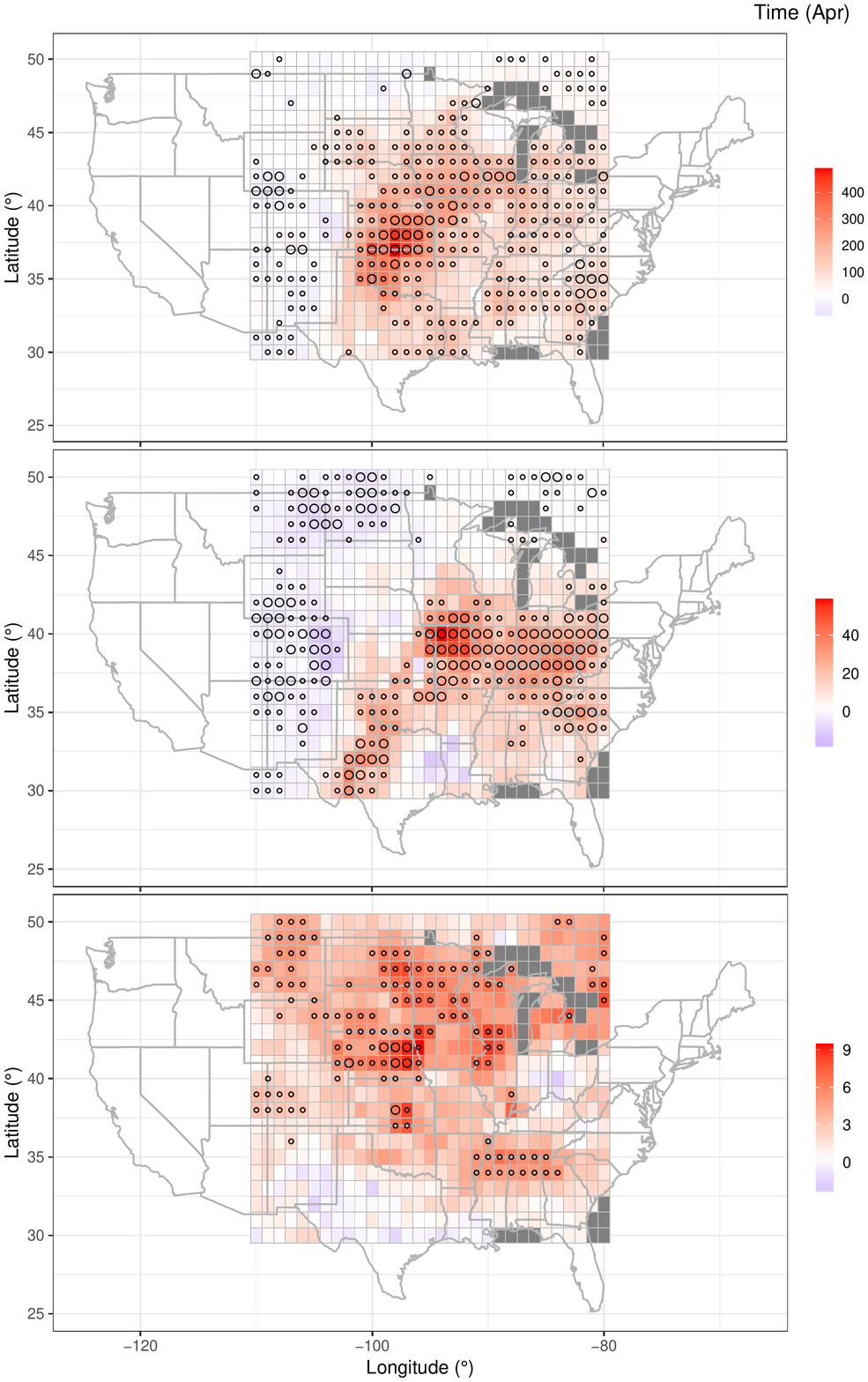}
\caption{Values and significance of the slope $\hat{\eta}_{1, \mathrm{ti}}$ for PROD (top), CAPE (middle) and SRH (bottom) maxima in April. 
Large and small circles indicate significance (after accounting for multiple testing using the BH procedure) at any level not lower than $5\%$ and $20\%$, respectively. The units of $\hat{\eta}_{1, \mathrm{ti}}$ are m$^3$s$^{-3}$yr$^{-1}$, Jkg$^{-1}$yr$^{-1}$ and m$^2$s$^{-2}$yr$^{-1}$ for PROD, CAPE and SRH, respectively. Dark grey corresponds to grid points where no observations are available.}
\label{Slope_Significance_Time_April}
\end{figure}

\begin{figure}
\center
\includegraphics[scale=0.67]{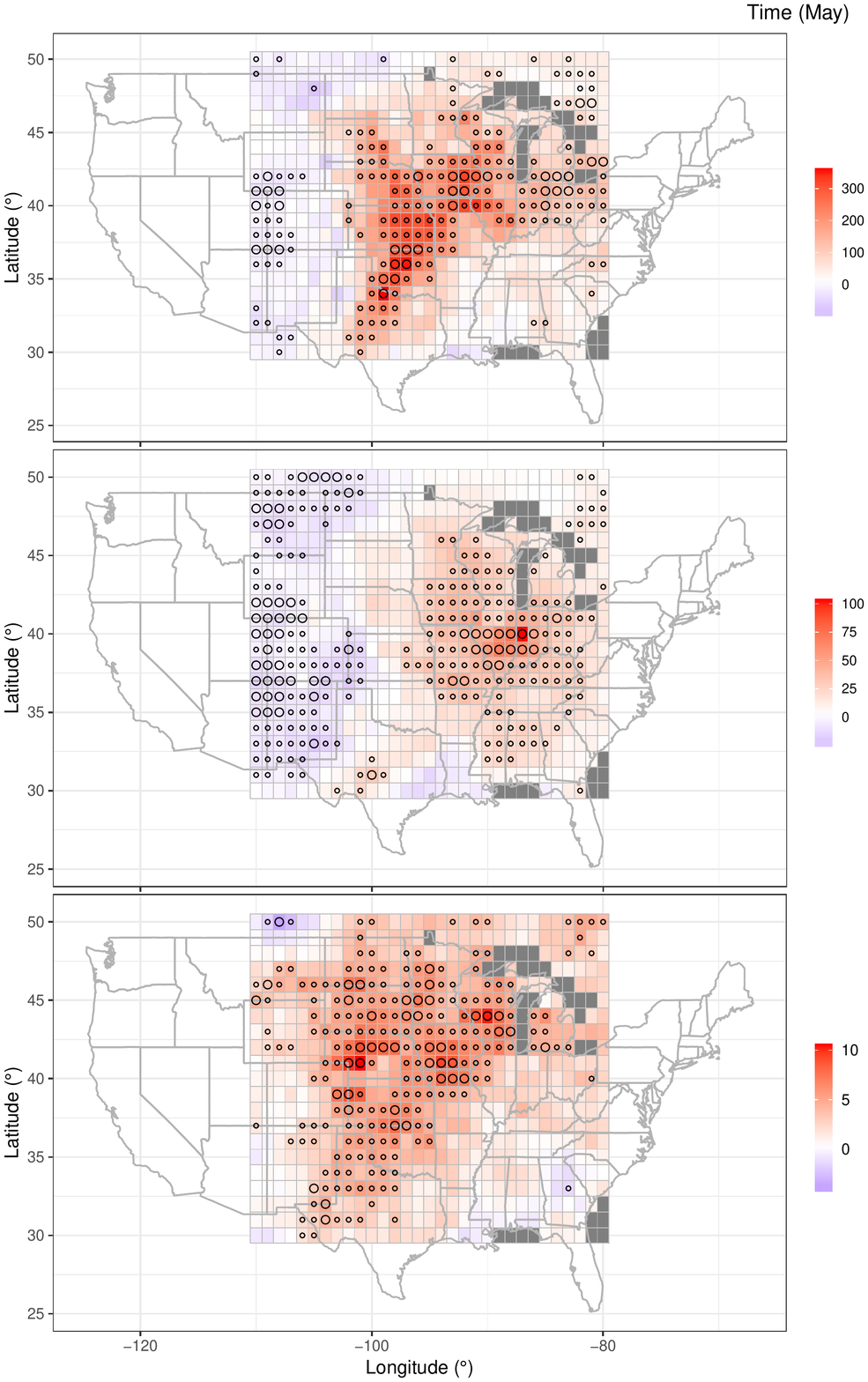}
\caption{Same content as in Figure~\ref{Slope_Significance_Time_April} in the case of May.} 
\label{Slope_Significance_Time_May}
\end{figure}

We begin with the effect of time. Table~\ref{Tab_Results_BH_All} shows that  many of the 619 grid points exhibit a significant time trend for PROD in April, May and August (and to a lesser extent in June and December). In April, this number equals $313$ at the $20 \%$ level, so~\eqref{Eq_First_Lower_Bound_Number_Significance} implies that at least $250$ of these grid points indeed have a trend; with~\eqref{Eq_Lower_Bound_Number_Significance}, this number rises to $278$. Figure~\ref{Slope_Significance_Time_April} indicates that, in April, the North-East, a very wide South-East corner and the South-West, show significant time trends. In the first two regions, $\hat{\eta}_{1, \mathrm{ti}}$ is positive, corresponding to an increasing risk of \Tb{severe thunderstorms impacts}, particularly in already risky regions. Similar conclusions may be drawn from Figure~\ref{Slope_Significance_Time_May} in the case of May, though the South-East is less prominent. \Tb{The highest slope value corresponds to an annual increase of PROD maxima of about $3\%$ of the corresponding PROD maximum recorded in $1979$.} \Tb{\cite{mannshardt2013extremes} and \cite{heaton2011spatio} do not find such a significantly positive time trend over the whole region which is the most at risk, sometimes called tornado alley, and they do not obtain significantly positive trends in the North-East of our region, whereas they find a significant positive trend towards the West. These differences are likely to be explained by the following facts: they consider a less recent period (1958--1999), their product variable is slightly different than ours, and they study annual instead of monthly maxima. The discrepancies with \cite{heaton2011spatio} can also be explicated by the methodological dissimilarities; as already mentioned, they use a Bayesian hierarchical approach. The evolution obtained by \cite{gilleland2013spatial} between the second (1979--1992) and the third (1993-1999) period is quite consistent with our trends in Spring; for the other seasons, however, the results differ significantly. There are also many dissimilarities  when considering the evolution between the first (1958--1978) and the second (1979--1992) periods; this comparison is less relevant since the first period does not belong to the time range we consider. \cite{gilleland2013spatial} consider the mean simulated values conditional on the total amount of energy being large, and then not all grid point values need be  extreme; on the other hand, we look at maxima at each grid point. Moreover, the trends we find account for the year-to-year variation, whereas in \cite{gilleland2013spatial}, changes can only be assessed by comparing  values for three successive periods of about 15 to 20 years. The positive time trends we obtain in Spring appear quite consistent with the results of \cite{Gensini2018}, who use much more recent data than the previously described papers. The remaining differences, especially for Texas, may arise for the following reasons. First, as PROD is only an indicator of severe weather, there are necessarily discrepancies with results based on effective tornado reports. Second, PROD slightly differs from STP, so the corresponding results may differ somewhat. Furthermore, the findings of \cite{Gensini2018} about reports concern the total number of tornadoes per year, and those about STP are not based on the maxima of that variable.}

Regarding CAPE, April, May and June (and to a lesser extent, August, November and January) show many grid points with a significant time trend. For April and May, Figures~\ref{Slope_Significance_Time_April} and~\ref{Slope_Significance_Time_May} show significantly negative $\hat{\eta}_{1, \mathrm{ti}}$ in the West, contrasting with a significantly positive trend in the center and the East. As pointed out by \cite{Trapp2009} and \cite{Diffenbaugh2013}, a positive time trend for CAPE is expected in a context of climate change. However, to the best of our knowledge, an \textit{observed} trend has not been previously reported in the literature.

For SRH, May and to a lesser extent April have many significantly positive grid points spread approximately uniformly except in a large South-West corner in April and a large South-East corner in May. The significance for PROD in April and May comes from both CAPE and SRH. Figures~\ref{Slope_Significance_Time_April} and~\ref{Slope_Significance_Time_May} suggest that the significant positive time trend in the \Tb{riskiest} part of the US stems mainly from CAPE in April and from SRH in May. \Tb{Overall, no seasonal pattern appears: CAPE seems to drive PROD in January, April, August, November and December, whereas SRH seems to drive it in February, May, June and September. For March, July and October, there is no clear driver. Anyway, trying to relate the behaviour of PROD maxima with that of CAPE and SRH maxima has its limitations. Indeed, the maximum of PROD may not equal the product of the square root of CAPE maximum and the maximum of SRH, as their maxima may not coincide.}

\begin{figure}
\center
\includegraphics[scale=0.67]{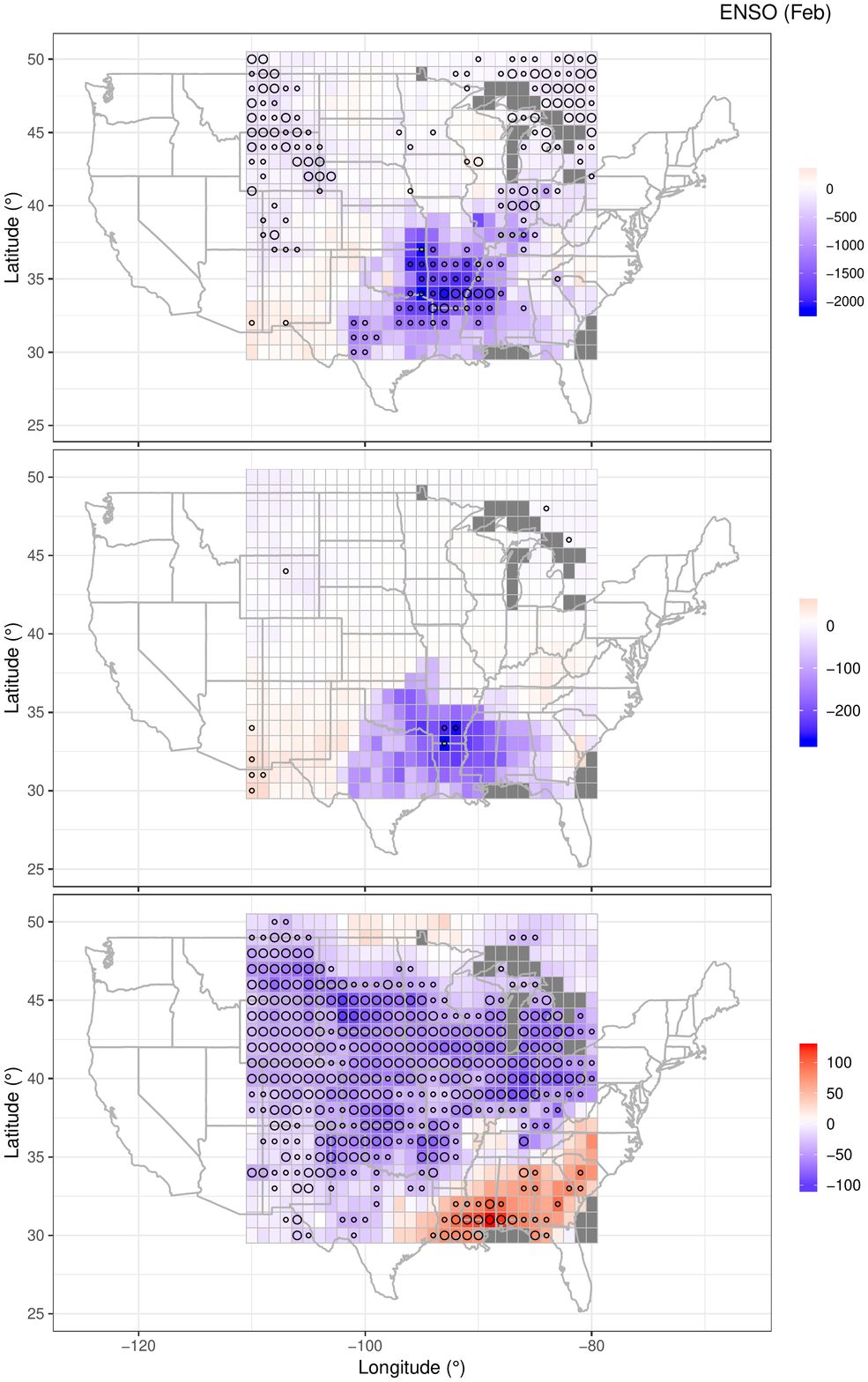}
\caption{Values and significance of \Tb{the ENSO coefficient} $\hat{\eta}_{1, \mathrm{en}}$ for PROD (top), CAPE (middle) and SRH (bottom) maxima in February. Large and small circles indicate significance (after accounting for multiple testing using the BH procedure) at any level not lower than $5\%$ and $20\%$, respectively. The units of $\hat{\eta}_{1, \mathrm{en}}$ are m$^3$s$^{-3}{}^{\circ}\mathrm{C}^{-1}$, Jkg$^{-1}{}^{\circ}\mathrm{C}^{-1}$ and m$^2$s$^{-2}{}^{\circ}\mathrm{C}^{-1}$ for PROD, CAPE and SRH, respectively.}
\label{Slope_Significance_ENSO_February}
\end{figure}

We now comment on the effect of ENSO. For PROD, Table~\ref{Tab_Results_BH_All} reveals that many grid points exhibit a significant relation in February. Figure~\ref{Slope_Significance_ENSO_February} indicates that $\hat{\eta}_{1, \mathrm{en}}$ is negative at those and that the main regions concerned are the North-East, the South-Center and the North-West; we expect higher PROD maxima during La Ni\~na years in these regions. \Tb{The highest slope absolute value corresponds to a decrease of PROD maxima per unit of ENSO of about $10\%$ of the corresponding basic level of PROD maximum.}

There is no strikingly significant  result for CAPE, although \cite{Allen:ENSO2014} found ENSO signals in CAPE seasonal averages for winter and spring, not accounting for multiple testing.

For SRH, a very large number of grid points exhibit significance in February. Figure~\ref{Slope_Significance_ENSO_February} shows that almost all grid points are concerned except for a strip in the North and a tiny diagonal strip in the South-East corner of the region. The estimate $\hat{\eta}_{1, \mathrm{en}}$ is highly negative in most of the region but very positive in the extreme South-East, with a very rapid change in sign, presumably due to proximity with the Gulf of Mexico. There is a significant negative relation in regions at risk of thunderstorms or large-scale storms, for which SRH plays an essential role. \Tb{The risk of large impacts} may increase during La Ni\~na years. A relationship between seasonal averages of SRH and ENSO in winter was noticed by \cite{Allen:ENSO2014}. Finally, Figure~\ref{Slope_Significance_ENSO_February} suggests that CAPE contributes more than SRH to PROD in terms of significance, though the relation with ENSO is more pronounced for SRH than for CAPE.

\begin{table}
\center
\renewcommand{\arraystretch}{1.5}
\resizebox{\textwidth}{!}{
\begin{tabular}{ccc||cccccccccccc}
\textbf{Variable} & \textbf{Covariate} &   \textbf{q}  &\textbf{Jan} &\textbf{Feb} &\textbf{Mar} &\textbf{Apr} &\textbf{May} &\textbf{Jun} &\textbf{Jul} &\textbf{Aug} &\textbf{Sep} &\textbf{Oct} &\textbf{Nov} &\textbf{Dec} \\ \hline   \hline
\textbf{PROD}  & \textbf{Time} & 0.05 &  7 &  0 &  1 & 41 & 36 &  0 &  0 & 36 &  2  & 0 &  0 & 22 \\ 
 & \textbf{Time} & 0.2 & 40 &  0 &  4 &313 &203 & 81 & 13 &148 & 23 &  0 &  0 & 98  \\ 
  & \textbf{ENSO} & 0.05 &  0 & 58 & 10 &  0 &  0 &  1 &  0 &  0 &  0 &  0 &  0 &  1 \\ 
  & \textbf{ENSO} & 0.2 &  1 &172 & 26 &  0 &  3 &  3 &  0 &  0 &  0 &  0 &  0 &  1 \\ 
\hline \hline
\textbf{CAPE} & \textbf{Time} & 0.05 & 37 & 13 & 28 &109 & 60 & 89 & 18 & 55 &  4 &  0 & 30 &  1  \\ 
   & \textbf{Time} & 0.2 & 92 & 37 & 73& 268 &273 &206 & 75 &133 & 35 & 40 &134 & 16  \\ 
  & \textbf{ENSO} & 0.05 & 15 &  0 &  0 &  0 &  0 &  2 &  2 &  0 &  0 &  0 &  1&   1  \\
  & \textbf{ENSO} & 0.2 & 27 & 11 & 21 &  0 &  0 &  3 & 16 & 14 &  0 &  1 &  6 & 13 \\ 
\hline \hline
\textbf{SRH} &   \textbf{Time} & 0.05 &  0 &  1 &  0 &  7 & 43 &  2 &  1 &  7 &  0 &  0 &  0 &  0 \\ 
  & \textbf{Time} & 0.2 & 15 & 44 &  4 &138 &230 & 14 & 50 & 45 &  6 &  0 &  0 & 27 \\ 
  & \textbf{ENSO} & 0.05 &  0& 255 &  0 &  0 &  1 &  0 &  0 &  0 &  0 &  0 &  0 &  0 \\ 
  & \textbf{ENSO} & 0.2  & 3 &384 & 59 & 18 &  4 &  0 &  8 &  7 &  4 &  1 &  0 & 82 \\ 
  \hline \hline
    & & & & & & & & & & & & & & \\  
    \hline \hline
\textbf{PROD res.} & \textbf{Time} & 0.05 &  7 &  0 &  2 & 30 & 88 &  0 &  0 & 41 &  2 &  0 &  0 & 38 \\
      & \textbf{Time} & 0.2 & 50 & 16  & 6 &274 &221 & 86 & 21 &137  &18  & 0 &  2 &100  \\ 
     \hline \hline
\textbf{CAPE res.} & \textbf{Time} & 0.05 & 35 & 20 & 15 & 87 & 96 & 89 & 25 & 59 &  9  & 0 & 19 &  2 \\
     & \textbf{Time} &  0.2 & 88 & 46 & 51& 219& 267 &223 & 91 &139 & 54 & 41 &120 & 29 \\ 
    \hline \hline
\textbf{SRH res.} & \textbf{Time} & 0.05 &  0 &  0 &  0 &  7 & 38 &  2 &  1 &  7 &  0 &  0 &  0 &  0   \\ 
    & \textbf{Time} &  0.2 & 20 &  1 &  6& 126& 241 &  7 & 46 & 41 &  1 &  0 &  0 & 60 \\ 
\hline \hline
    & & & & & & & & & & & & & & \\  
    \hline \hline
\textbf{PROD res.} & \textbf{ENSO} & 0.05 &  1 & 66 &  8 &  0 &  0 &  1 &  0 &  0 &  0 &  0 &  0 &  7 \\
				  & \textbf{ENSO} & 0.2  & 1  & 178 & 26 &  0 & 49 &  3 &  0 &  0 &  0 &  0 &  0 & 33 \\
				  \hline \hline
\textbf{CAPE res.} & \textbf{ENSO} & 0.05 &  1 &  0 &  0 &  0 &  0 &  3 &  5 &  1 &  0 &  0 &  0 &  2 \\
				  & \textbf{ENSO} & 0.2  & 21 & 38 &  0 &  1 &  0 &  4 & 17 & 16 &  0 &  0 &  1 & 21 \\
				  				  \hline \hline
\textbf{SRH res.}  & \textbf{ENSO} &  0.05 &  0 & 209 & 0 &  0 &  4 &  0 &  0 &  0 &  1 &  0 &  0 &  0 \\
 				  & \textbf{ENSO} & 0.2 &  1 & 359 & 20 & 38 & 14 &  0 &  3 &  7 &  2 &  1 &  0 & 63 \\
\end{tabular}
}
\caption{Number of grid points where $\hat{\eta}_{1, \mathrm{ti}}$ and $\hat{\eta}_{1, \mathrm{en}}$ are significant for PROD, CAPE and SRH maxima for each month (top); number of grid points where $\hat{\eta}_{1, \mathrm{ti}}$ is significant for PROD, CAPE and SRH maxima residuals after accounting for the relation with ENSO (middle); number of grid points where $\hat{\eta}_{1, \mathrm{en}}$ is significant for PROD, CAPE and SRH maxima residuals after accounting for the relation with time (bottom). We have accounted for multiple testing using the BH procedure with the values of $q$ displayed.}
\label{Tab_Results_BH_All}
\end{table}

We also considered the residuals of PROD, CAPE and SRH maxima after accounting for ENSO or temporal effects. For instance, if we observe a time trend, the idea of considering the residuals after accounting for ENSO is to determine whether the time trend is explained by ENSO. This allows us to determine whether the time and ENSO effects are ``independent''. 

In the case of PROD, Table~\ref{Tab_Results_BH_All} shows that removing ENSO does not much decrease the number of grid points exhibiting a significant time trend; there is a slight decrease for April but a small increase for some other months. Accounting for the time trend, on the other hand, can slightly increase the number of grid points showing a significant relation with ENSO.  For CAPE, removing ENSO decreases the number of grid points exhibiting a significant time trend for March, but there is a slight increase for other months, whereas accounting for time slightly decreases the number of grid points showing a significant relation with ENSO in January and March only, with a  slight increase in other months. Regarding SRH, removing ENSO decreases the number of grid points exhibiting a significant time trend in February but there is little impact for other months. The conclusions are similar when accounting for the time trend and studying the ENSO effect. The maps of the residuals (not shown) indicate that when removing a covariate has little impact on the number of grid points at which the relation with the other covariate is significant, it has almost no impact on their positions either. To summarize, the effects of time and ENSO appear ``independent'', except for CAPE in January and March and SRH in February. 

\section{Conclusion}
\label{Sec_Conclusion}

This article quantifies the effects of time and ENSO on the distribution of monthly maxima of PROD, CAPE and SRH, which are highly relevant to the risk of severe thunderstorms. The use of the GEV appears  justified in our setting. After allowance for multiple testing we detect a significant time trend in the location parameter of the GEV for PROD maxima in April, May and August, CAPE maxima in April, May and June and SRH maxima in April and May. The observed upward time trend for CAPE, although expected in a warming climate, has not been reported before. April and May are prominent for PROD, as severe thunderstorms are common at this period, and the corresponding trend is positive in parts of the US where the risk is already high, which may have important consequences. We also found ENSO to be a good covariate in the location parameter of the GEV for PROD and SRH maxima in February. The corresponding relationship is negative over most of the region we consider, suggesting that \Tb{the risk of storm impacts} in February increases during La Ni\~na years. \Tb{Our results differ from those of \citet{heaton2011spatio}, \cite{mannshardt2013extremes} and \cite{gilleland2013spatial}, but are quite consistent with those obtained by \cite{Gensini2018}, perhaps in part because these authors consider a period similar to ours, more recent than in the earlier studies.} 

We investigate the effects of time and ENSO on the marginal (at each grid point) extremal behaviour of PROD, CAPE and SRH. Quantifying the potential impacts of these covariates on the local spatial extremal dependence of these variables would also be useful for risk assessment. \Tb{Modelling} the extremal dependence between CAPE and SRH might also be informative.

\Tb{An interesting question is the implication of an increase of PROD (or SRH) maxima. As PROD can be seen as a proxy for the probability of severe thunderstorm occurrence, it is natural to think that PROD maxima may be good indicators for the maxima of the variable ``number of severe thunderstorms per day". This would somehow imply that the days where PROD maxima occur generally correspond to those days with the largest severe thunderstorms impacts. Providing clear insight about whether this is indeed the case would be valuable.} 

\section*{Acknowledgements}

The work was supported by the Swiss National Science Foundation (project 200021\_178824). NARR data were downloaded from the Research Data Archive (RDA) at the National Center for Atmospheric Research (NCAR), Computational and Information Systems Laboratory at \url{http://rda.ucar.edu/datasets/ds608.0/}. The ERSSTv5 from the NOAA Climate Prediction Center is available at \url{https://www.cpc.ncep.noaa.gov/data/indices/ersst5.nino.mth.81-10.ascii}. 

\newpage
\bibliographystyle{apalike}
\bibliography{Biblio_Final}

\end{document}